%% file: main.tex
\documentclass[acmsmall]{acmart}
\authorsaddresses{}
\usepackage{booktabs}
\usepackage{multirow}
\usepackage{graphicx}
\usepackage{pgfplots}
\pgfplotsset{compat=1.18}
\usepackage{tikz}
\usetikzlibrary{arrows.meta,positioning,shapes.geometric,fit,backgrounds,calc}
\usepackage{xcolor}
\usepackage{arydshln}
\usepackage{balance}
\usepackage{subcaption}
\usepackage{amsmath}
\usepackage{url}
\usepackage{wrapfig}
\usepackage{hyperref}
\usepackage[normalem]{ulem} 
\usepackage{algorithm}
\usepackage{algpseudocode}
\usepackage{array}
\newcommand{\para}[1]{\par\vskip\smallskipamount\noindent\textbf{#1}\enspace}

\microtypesetup{expansion=false}

\definecolor{classicalcol}{RGB}{31,119,180}
\definecolor{pqccol}{RGB}{255,127,14}
\definecolor{compositecol}{RGB}{44,160,44}
\definecolor{hybridcol}{RGB}{214,39,40}
\definecolor{collapsecolor}{RGB}{255,230,230}
\definecolor{tableheader}{RGB}{30,55,95}
\usepackage{colortbl}
\usepackage{enumitem}

\newcommand{\circnum}[1]{%
  \tikz[baseline=(n.base)]{%
    \node[circle, fill=black, inner sep=0.5pt, minimum size=9pt] (n)
      {\textcolor{white}{\scriptsize\bfseries#1}};%
  }%
}

\newcommand{\hide}[1] {}
\newcommand{\sknote}[1]{{{\textcolor{magenta}{SK:{#1}}}}}

\AtBeginDocument{%
  }

\begin{document}

\title{PQC Migration of IPsec: Reliable Composite Key Sharing Mechanisms in IKEv2}

\author{Rahul}
\affiliation{%
  \institution{Indian Institute of Technology Bombay}
  \city{Mumbai}
  \country{India}}
\email{iamrahul823@gmail.com}

\author{Manjesh K. Hanawal}
\affiliation{%
  \institution{Indian Institute of Technology Bombay}
  \country{India}}

\author{Sameer G. Kulkarni}
\affiliation{%
  \institution{Indian Institute of Technology Gandhinagar}
  \country{India}}

  \author{Tirumaleswar Reddy K}
\affiliation{%
  \institution{Nokia}
  \country{India}}


\begin{abstract}
\input{sections/abstract}

\end{abstract}

\hide{

\ccsdesc[500]{Security and privacy~Security protocols}
\ccsdesc[300]{Security and privacy~Mobile and wireless security}
\ccsdesc[100]{Networks~Routing protocols}

\keywords{Post-quantum cryptography, IKEv2, IPsec, ML-KEM, TCP transport,vcomposite key exchange, energy measurement, RAPL, strongSwan}
}
\maketitle
\input{sections/01_intro}

\input{sections/02_background}
\input{sections/03_implementation}

\input{sections/04_setup}
\input{sections/05_algorithms}
\input{sections/06_results}
\input{sections/08_related}
\input{sections/09_conclusion}

\bibliographystyle{unsrt}
\bibliography{references}
\input{sections/appendix.tex}

\end{document}

%% file: sections/abstract.tex
The threat of quantum computers breaking classical cryptographic is accelerating the adoption of quantum-resistant cryptographic algorithms. Several government agencies have set a tight time-line for upgradation of all system to support post quantum cryptographs. 
We study post-quantum migration of IPSec protocols and their impact on network performance. The large key lengths in Post-Quantum Cryptography (PQC) cannot fit in a single IP packet leading to IP fragmentation. This can cause failure of security associations and degrade performance of IKEv2 under poor network conditions. 
To tackle this problem, we study a new key-sharing mechanism for IKEv2 and evaluate its performance. Unlike current hybrid key-sharing approaches, which send the classical and quantum keys sequentially over UDP or TCP (as in RFC 9370), the new approach concatenates them into a single \textit{composite key} and transport them over TCP to ensure a reliable handshake. 



We implement and benchmark the new composite key sharing handshake mechanism against the existing hybrid algorithms that use either UDP or TCP. Extensive experiments under various network conditions demonstrate that the new approach significantly improves handshake time and energy consumption by avoiding the severe retransmission penalties of the Hybrid handshake mechanism. At 1\% packet loss, the handshake time of composite key sharing mechanism is better by atleast 20\% than hybrid mechanisms, and the improvement get better with higher packet loss reaching 30\%  at 10\% packet loss. Energy consumption follows the same pattern achieving reduction of 29\% at 10\% packet loss.

%% file: sections/01_intro.tex
\section{Introduction }
\label{sec:intro}
Every time a VPN connection is established or a corporate network is accessed remotely, protocols such as Internt Protocol Security (IPsec) and Internet Key Exchange verion 2 (IKEv2) work in the background to secure the communication. They authenticate peers, negotiate cryptographic keys, and create encrypted tunnels to protect data as it travels across public networks. 
The security of these systems relies on the computational complexity of reversing cryptographic algorithms to recover the keys. However, advances in quantum computing (QC) invalidate these assumptions to pose a significant threat to the security of VPN connections worldwide.  
Post-quantum cryptographic (PQC) algorithms address this emerging threat by preserving the cryptographic security through the mathematical constructs that are believed to remain computationally intractable even for quantum computers~\cite{ncsc_pqc_timeline,nistpqc_migration}. This resilience is achieved through the use of quantum-resistant mathematical constructions (\textit{e.g.}, lattices, hash functions, or multivariate polynomials), which often require large key sizes than those used in classical cryptographic algorithms~\cite{chen2016report}.
Several government agencies have announced a roadmap for migration to post-quantum cryptographic algorithms in security protocols, and some have set a deadline for migrations as early as 2031 \cite{ncsc_pqc_timeline}\cite{nistpqc_migration}.
NIST's August 2024 standardization of ML-KEM~\cite{fips203} resolved the cryptographic question of which key encapsulation mechanism replaces Diffie-Hellman. 
However, for the operators running IKEv2~\cite{rfc7296} gateways, it opened a different question, \textit{i.e., how can ML-KEM be deployed while maintaining the scalability and performance required by production VPN infrastructures?}

With Curve25519, one of the most widely used Elliptic Curve Cryptography (ECC), the 256-bit (32-byte) key results in a  262-byte \textsc{ike\_sa\_init} message 
that can easily fit within a single UDP datagram on any standard Ethernet network (MTU\,=\,1,500\,bytes).
In contrast, with ML-KEM-768, the key size increases to 1,184 bytes, resulting in a payload message of nearly 1,400 bytes \textsc{ike\_sa\_init} message.  
Moreover, with ML-KEM-1024~\cite{fips203} the key size increases to 1,568 bytes and correspondingly the \textsc{ike\_sa\_init} message size increases to $\sim$1800 bytes, exceeding the standard MTU and forcing it to be fragmented. 
\hide{The cryptography is sound, but the transport is broken!}
Consequently, with PQC, although the cryptography is strengthened, the underlying transport becomes more fragile, exposing VPN deployments to performance degradation issues, such as, increased latency, retransmissions, and lower connection establishment performance.
Moreover, the transition to PQC cannot be achieved by simply replacing classical algorithms, as the established classical public-key schemes have undergone decades of cryptanalysis and are well understood against classical adversaries, Whereas PQC algorithms, despite extensive evaluation and standardization, have comparatively shorter security histories.
This demands to leverage the strengths of classical algorithms alongside PQC during the transitional period. Hence, the current deployment advocates the use of \textit{hybrid key-sharing} in which both classical and post quantum crypto algorithms are used together. This ensures the resulting session key to remain secure, provided at least one of the constituent schemes is not broken. However, this approach further increases the handshake message size by transmitting both classical and post-quantum keys to further worsen the IP fragmentation issues and the transport burden.

To address the fragmentation problem, RFC~7383~\cite{rfc7383} introduced `IKE Fragmentation' that splits the oversized IKE payloads at the IKE protocol layer before they reach IP, enabling per-fragment retransmission to avoid IP-layer reassembly failures. However, this did not solve the IP fragmentation of the crypto payload as IKE fragmentation cannot\footnote{RFC~7383 was designed primarily to handle IP fragmentation caused by large authentication certificates in \textsc{ike\_auth}.}$^,$\footnote{IKE fragmentation is designed to fragment only encrypted IKE messages because encryption along with integrity protection ensure that the receiver can safely reassemble the fragments without exposing the protocol to new attacks.} be used in the initial IKEv2 exchange \textsc{ike\_sa\_init}. 
RFC~9242~\cite{rfc9242} defined 'Intermediate Exchange' or \textsc{ike\_intermediate}as a new exchange for IKEv2 where one or more message exchanges may take place right after the \textsc{ike\_sa\_init}. 
This enables IKE messages to be fragmented at the application layer, allowing payloads exceeding the path MTU to be transmitted without IP fragmentation.
RFC~9730~\cite{rfc9730} leverages this mechanism to realize hybrid key exchange, in which the \textsc{ike\_sa\_init} message carries the classical key share, and subsequent 
\textsc{ike\_intermediate} messages carry the PQC key share.
All the IKE messages exchanged during this hybrid key-sharing use UDP. 
Although functional, the current hybrid key-sharing with UDP poses three fundamental limitations that hinder its practical deployment: 
\newline\noindent
\textbf{Performance degradations:} The larger handshake messages in hybrid post-quantum key exchange increase the likelihood of IP fragmentation, making connection establishment substantially more sensitive to packet loss as the loss of a single fragment forces retransmission of the entire message. In IKEv2, such losses trigger application-layer retransmission timers that typically operate in a 1-2s timeout~\cite{rfc7296}. This leads to substantial increase in handshake latency even under modest network loss conditions. Our benchmarks show that even $ \leq 1\%$ loss-rate can result in 2-4$\times$ increase in the connection establishment time.
    \newline\noindent
    \textbf{Security:} PQ crypto payload exchanged in \textsc{ike\_intermediate} must still be protected using the keys derived from the classical key exchange \textsc{ike\_sa\_init} message to ensure reliable delivery and reassembly over UDP. Hence, the security of the overall mechanism remains partially anchored in classical cryptography, rather than being fully independent of it.
    \newline\noindent
    \textbf{PQ-only support}: as cryptographically relevant quantum computers (CRQCs) become more prevalent, the regulations are likely to mandate the use of post-quantum only systems. However, the current  hybrid key-sharing over UDP cannot be used in such a scenario as it relies on classic crypto to handle IP fragmentation issues. 
To overcome these limitations, we design, prototype and evaluate a novel composite hybrid key exchange mechanism proposed in an IETF Internet-Draft~\cite{reddydraft}.
Instead of exchanging classical and post-quantum key shares over multiple round trips, the proposed mechanism combines both components into a single composite public key.
It builds upon the \textsc{separate\_transports}~\cite{ietf_reliable_transport} which decouples the transport of control and data traffic by carrying IKEv2 control messages over TCP while preserving UDP transport for ESP~\cite{rfc4303} packets.
This ensures to reliably transport the large \textsc{ike\_sa\_init} messages over TCP using the encapsulation format defined in RFC~9329~\cite{rfc9329}.
We evaluate the performance of this novel composite key-sharing mechanism of IKEv2 with TCP. 
Even though, the TCP was recommended for key-exchange in RFC~9329, it did not find wider adaptation as the RFC~9329 maintained TCP for subsequent ESP messages without any transport switch\footnote{We note that the RFC 9329 addressed a situation where UDP port is blocked with only TCP left for message exchanges}. This resulted in nested TCP resulting degradation of performance due to TCP meltdown~\cite{honda2005tcp}. The new composite mechanism uses TCP for IKEv2 key exchange and UDP for ESP payloads through a transport switch.  Our evaluation supports that this transport switch does not degrade the performance while offering a reliable transport needed for longer PQ keys. We believe, our extensive evaluation support PQC migration of IPsec with the composite key-sharing mechanism while maintaining the scalability and performance requirements of VPN infrastructure.  

We implement the composite key-sharing algorithm 
on strongSwan~6.0~\cite{strongSwan}. 
\hide{The composite KE plugin registers Transform Type~4 identifiers for each classical/PQC pair and serializes both key shares into a single \textsc{ike\_sa\_init} payload by concatenation.}
For reproducibility, a local fork with our modifications is made available at
~\cite{anon_repo}.
\hide{ \sknote{All of this should go in Evaluation Methodology only!} We evaluate the performance of the composite key-sharing algorithm with a combination of three classical  (Curve25519, ECP-256, ECP-384) and three post-quantum  (ML-KEM-512, ML-KEM-768, ML-KEM-1024~\cite{fips203}) crypto suites. Specifically, we focus on three composite key-shares, namely Curve25519$\Vert$ML-KEM-768, ECP-256$\Vert$ML-KEM-768, ECP-384$\Vert$ML-KEM-1024, where  $\Vert$ indicates the concatenation operation. 
The combination of classic and post-quantum crypto suites are recommended as good practical options in ~\cite{reddydraft}. 
We pair each of the composite key-shares with three post-quantum Digital Signature Algorithms (DSA): ML-DSA-44, ML-DSA-65, and ML-DSA-87~\cite{fips204}. We compare the performance of each composite key-share across several metrics, including handshake latency, throughput, and energy consumed for key exchange, under a controlled environment.
} 
We emulate realistic network conditions by varying both packet loss rates and network delay parameters.
Our evaluation suggest that the composite key-share over TCP outperforms the hybrid key-share, with increasingly larger gains under poor network conditions. 
In summary, our contributions include:
\begin{itemize}
\item We implement novel 
composite key-sharing mechanism 
on IKEv2 stack of strongSwan~6.0~\cite{strongSwan} and verify the operational correctness of separate transports 
for different key pairs (\S\ref{sec:impl}).

\item We build an experimental testbed to benchmark composite and hybrid key-sharing algorithms over TCP and UDP, evaluating handshake latency, throughput, and energy consumption under varying packet loss and network delay conditions (\S\ref{sec:eval}).


\item Our evaluation demonstrates that composite key-sharing over TCP consistently outperforms hybrid key-sharing across all evaluated network conditions. Furthermore, the performance benefits increase with deteriorating network conditions. 
Our findings establish composite key-sharing over TCP to be a more reliable mechanism for PQC in IPSec (\S\ref{sec:results}). 


\end{itemize}

%% file: sections/02_background.tex
\section{Background}
\label{sec:background}
In this section, we discuss the structure of IKEv2 to support hybrid key sharing over UDP and TCP, and the need for a composite key-sharing mechanism via reliable transport. 

\subsection{IKEv2 Exchange Structure}

IPsec~\cite{rfc4301} secures traffic at the IP layer through two complementary
protocols: IKEv2~\cite{rfc7296} for the control plane and Encapsulating Security Payload (ESP)~\cite{rfc4303} for the data plane. 
Before any user traffic is protected, the two peers complete an IKEv2 handshake to negotiate cryptographic
algorithms, authenticates identities, and derives the session keys that protect subsequent ESP traffic. 
This IKEv2 handshake proceeds in two mandatory
exchanges, namely, 
(i) \textsc{ike\_sa\_init} where it carries Diffie-Hellman public keys, nonces, and the proposed algorithm suite. The output is a shared secret from which IKE Security Association keys are derived. 
(ii) \textsc{ike\_auth} follows
immediately. It uses those keys to authenticate peer identities via digital signatures (or pre-shared-key proofs) and establishes the first Child SA that ESP uses for encrypted user traffic.

When both peers support post-quantum hybrid key exchange, 
one or more \textsc{ike\_intermediate} exchanges
are inserted between \textsc{ike\_sa\-\_init} and \textsc{ike\_auth}~\cite{rfc9370}. 
Each carries an Additional Key Exchange (ADDKE) payload containing the post-quantum algorithm's public key or ciphertext. 
A single ML-KEM-768 ADDKE payload is 1088--1184 bytes, when sent over UDP, it may need to undergo fragmentation at the IP layer or be handled by IKEv2 message fragmentation~\cite{rfc7383}. 

Further, RFC~7383~\cite{rfc7383} defines application-layer fragmentation for IKEv2 where oversized IKE messages are split into \textsc{IKE\_FRAGMENT} datagrams before transmission, with each fragment sent as a separate UDP datagram. This avoids IP-layer reassembly problems and enables per-fragment retransmission. However, it does not eliminate UDP loss as each fragment remains an independent datagram. Thus, any single fragment loss triggers the
IKEv2 retransmission timer~\cite{rfc7296} just as a full-message loss would. Under packet loss, RFC~7383 reduces worst-case stall duration but cannot prevent the timer from firing.
The round-trip count matters independently of payload size. Classical IKEv2~\cite{rfc7296} requires two round trips
(\textsc{ike\_sa\_init} and \textsc{ike\_auth}). Adding one
\textsc{ike\_intermediate} for RFC~9370~\cite{rfc9370} Hybrid key exchange requires three. Each extra round trip is an additional opportunity for a loss event to restart the 1.0-second base timer, and under high loss rates
that difference compounds quickly.

\subsection{Composite Key exchange in IKEv2}
\label{sec:background:draft}

The Composite key exchange~\cite{reddydraft} (currently an IETF draft) addresses two structural problems with the RFC~9370~\cite{rfc9370} approach, namely, i) the extra round trip introduced by \textsc{ike\_intermediate}, and ii) the IP fragmentation exposure arising from large PQC payloads over UDP. 
Accordingly, it defines three coordinated mechanisms to eliminate them.

First, it registers three Transform Type~4 (Key Exchange Method) identifiers, each treating a pair of classical and post-quantum key material as a single, indivisible negotiation unit. Currently. the three identifiers are named as (\texttt{TBD1: ecp256+mlkem768}), (\texttt{TBD2: ecp384+mlkem1024}), and (\texttt{TBD3: curve25519+mlkem768}). 
A responder either accepts or rejects the the full composite pair in its entirety with no provision for partial acceptance. As this operation maps to a single Transform ID, the need for separate \textsc{ike\_intermediate} or Additional Key Exchange (ADDKE) is eliminated, which consequently reduces the IKEv2 handshake from three round trips to two. 
\noindent\newline\textbf{Combined key exchange payload}: Both peers serialize their key material by concatenation. The initiator sends
$\mathit{KE}_i = \langle k_\text{classical} \Vert k_\text{PQ} \rangle$,
and the responder sends $\mathit{KE}_r = \langle k_\text{classical} \Vert
c_\text{PQ} \rangle$, where $k$ denotes a public key and $c_\text{PQ}$ is the ML-KEM ciphertext. The byte boundary between components is fixed by the Transform ID as the classical and PQC key sizes are constants for each
algorithm pair.
This results in \textsc{ike\_sa\_init} message size  ranging from 1,456--1,888 bytes, exceeding the Ethernet MTU and increasing the likelihood of IP fragmentation when carried over UDP. As a result, it makes TCP transport a protocol requirement rather than an optimization.
\newline\noindent\textbf{Reliable TCP transport for composite key-share}:
Second, the draft resolves the transport question using the split-plane configuration as shown in 
Figure~\ref{fig:splitplane}.
When the initiator begins \textsc{ike\_sa\_init} over a TCP connection per RFC~9329~\cite{rfc9329} and includes a \textsc{separate\_transports} notify
payload (type~16460), and the responder echoes it, both peers commit to TCP connection for all IKEv2~\cite{rfc7296} control-plane messages, including rekeying and liveness checks, while the ESP~\cite{rfc4303} data-plane packets continue over UDP.

TCP performs segment-level retransmission at transport layer, recovering the lost segments transparently and typically on a much shorter timescale than IKEv2's message retransmission timers, making packet loss effectively transparent to the IKEv2 protocol. 
In contrast, ESP on UDP preserves the data plane's latency characteristics and avoids the nested-TCP congestion-control instability that arises when both control and data planes share a single TCP connection, a phenomenon referred as TCP meltdown ~\cite{tcpmeltdown}.

Fig.~\ref{fig:exchange} illustrates the structural difference between Hybrid and Composite key exchanges. 
As shown in Fig.~\ref{fig:exchange}(a) and ~\ref{fig:exchange}(b), shown the Hybrid over TCP and Hybrid over UDP, respectively. The Hybrid path inserts an \textsc{ike\_intermediate} carrying the ML-KEM ciphertext between the two standard exchanges over UDP or TCP. 

\input{figures/fig_splitplane}
Each of the three round trips provide an independent opportunity for fragment loss, which can stall the handshake.
On the other hand, the Composite TCP shown in Fig.~\ref{fig:exchange}{c},  sends both classical and post-quantum keys in a single \textsc{ike\_sa\_init} over TCP, completing in two round-trip with TCP's segment-level recovery invisible to the IKEv2.
%
%

%
\noindent\textbf{Key Derivation in Composite key-share}:
Once both peers exchange the combined KE payload, each independently
derives a key $K_T$ from classical ECDH and a key $K_{PQ}$ from
ML-KEM decapsulation~\cite{fips203}. Both the key are combined using the Hash-based Message Authentication Code (HMAC)-based Key  Derivation Function (HKDF)-based Universal Combiner as described in the 
hybrid-KEM document~\cite{rfc5869, hybridkems}:
\begin{equation}
  K_{\text{combined}} = \mathrm{UC}\!\left(K_{PQ},\,K_T,\,\mathit{CT}_{PQ},\,
  \mathit{PK}_{PQ},\,\mathit{PK}_T,\,\texttt{"IKEv2-HYBRID-KE-v1"}\right)
\end{equation}
where $\mathit{CT}_{PQ}$ is the ML-KEM ciphertext and $\mathit{PK}_{PQ}$,
$\mathit{PK}_T$ are the post-quantum and classical public keys respectively.
The domain-separation label \texttt{"IKEv2-HYBRID-KE-v1"} binds the output
to this specific protocol and version, preventing cross-protocol key
confusion even if the same key material appears in another context.

The combined key replaces the classical $g^{ir}$ term in the standard
RFC~7296~\cite{rfc7296} key schedule:
\begin{equation}
  \mathrm{SKEYSEED} = \mathrm{prf}(N_i \mathbin{\Vert} N_r,\; K_{\text{combined}})
\end{equation}
where $N_i$ and $N_r$ are the initiator and responder nonces from
\textsc{ike\_sa\_init}. All subsequent IKE keying material
are derived from SKEYSEED exactly as specified in RFC~7296~\cite{rfc7296}, so the rest of
the key schedule requires no modification. The post-quantum security of the
combined key holds as long as either $K_T$ or $K_{PQ}$ remains
computationally infeasible to recover. 

\input{figures/HandShakeVariants}

%% file: figures/fig_splitplane.tex
\begin{wrapfigure}{r}{0.35\columnwidth}
\centering
\scalebox{0.60}{%
\begin{tikzpicture}[
  font=\scriptsize,
  >=Stealth,
  ep/.style ={rectangle, draw=gray!55, fill=gray!8, thick,
              minimum width=1.55cm, minimum height=0.56cm, align=center},
  mid/.style={rectangle, draw=#1!70, fill=#1!10, thick,
              minimum width=1.85cm, minimum height=0.56cm,
              align=center, font=\scriptsize\bfseries},
]

\draw[rounded corners=3pt, draw=gray!40, thick, fill=gray!3]
  (0, 0.05) rectangle (7.2, 1.65);
\node[anchor=north, font=\scriptsize\bfseries] at (3.6, 1.62)
  {Control Plane (IKEv2)};

\node[ep]               (ci) at (0.95, 0.75)  {Initiator};
\node[mid=classicalcol] (ct) at (3.60, 0.75)  {TCP port 4500};
\node[ep]               (cr) at (6.25, 0.75)  {Responder};

\draw[<->, classicalcol!80, thick, >=Stealth]
  (ci.east) -- (ct.west);
\draw[<->, classicalcol!80, thick, >=Stealth]
  (ct.east) -- (cr.west);

\draw[rounded corners=3pt, draw=gray!40, thick, fill=gray!3]
  (0, -1.75) rectangle (7.2, -0.15);
\node[anchor=north, font=\scriptsize\bfseries] at (3.6, -0.18)
  {Data Plane (ESP)};

\node[ep]         (di) at (0.95, -0.95)  {Initiator};
\node[mid=pqccol] (ut) at (3.60, -0.95)  {UDP port 4500};
\node[ep]         (dr) at (6.25, -0.95)  {Responder};

\draw[<->, pqccol!80, thick, >=Stealth]
  (di.east) -- (ut.west);
\draw[<->, pqccol!80, thick, >=Stealth]
  (ut.east) -- (dr.west);

\node[font=\scriptsize, compositecol, align=center] at (3.6, -2.05)
  {\textsc{separate\_transports}: IKEv2 $\to$ TCP \quad ESP $\to$ UDP};

\end{tikzpicture}
}%
\caption{Split-plane transport negotiated by \textsc{separate\_transports}: IKEv2 uses TCP and ESP uses UDP.}
\label{fig:splitplane}
\end{wrapfigure}
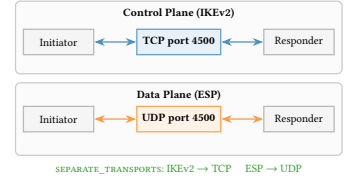

%% file: figures/HandShakeVariants.tex
\begin{figure*}[!h]
\centering
\includegraphics[width=\textwidth]{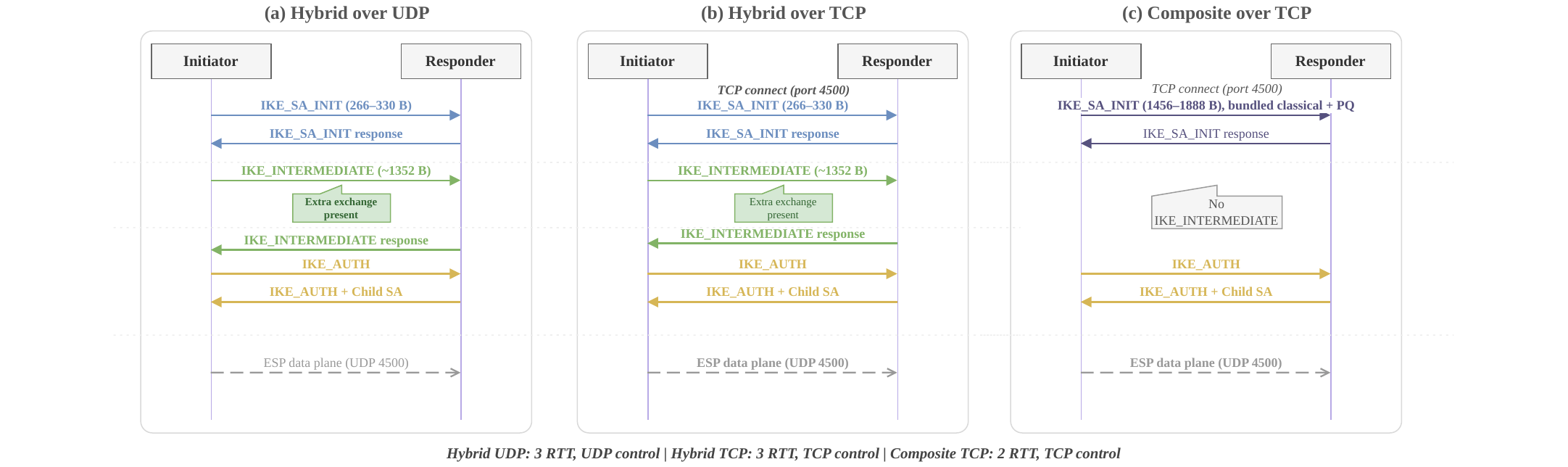}
\caption{IKEv2 handshake variants: Hybrid UDP vs Hybrid TCP vs Composite TCP}
\label{fig:exchange}
\end{figure*}

%% file: sections/03_implementation.tex
\section{Implementation: Composite key-share support in strongSwan}
\label{sec:impl}

To the best of our knowledge, no implementation of the proposed composite key exchange is currently available. 
We therefore implement it in the widely used open-source IPsec implementation strongSwan.
Our prototype is built on the \texttt{ml-dsa} development branch of strongSwan~6.0~\cite{strongSwan}, and compiled from source against liboqs~0.8.0 and oqs-provider~0.5.0 and OpenSSL~3.x. 
We extend the \texttt{charon} IKEv2 daemon by implementing two independently loadable plugins the \texttt{composite-ke} plugin and the \texttt{tcp-socket} plugin  to facilitate the composite key exchange over TCP.
The following sections describe the new additions and updates to strongSwan to support composite key-sharing. 
This updated version of strongSwan (currently anonymized) is made available at~\cite{anon_repo}.


\noindent\textbf{Composite KE plugin}: This plugin registers three Transform Type~4 (Key Exchange Method)
identifiers as defined in the  draft~\cite{reddydraft}, namely i) \texttt{ ecp256+mlkem768}, ii) \texttt{ecp384+mlkem1024}, and iii) \texttt{curve25519+mlkem768}. 
We perform payload serialization via direct concatenation without separators or length prefixes between components. Note that no explicit delimiter is used, as byte boundaries are determined solely by the fixed key sizes implied by the transform identifier.

For the key combination of \texttt{ curve25519+mlkem768}, the initiator writes a 32-byte X25519 public key immediately followed by the 1,184-byte ML-KEM-768 public key~\cite{fips203} into a
single \textsc{ke} payload buffer. 
The responder reads at the same fixed offsets, performs X25519 scalar multiplication and ML-KEM-768 decapsulation independently, then serializes the response as the 32-byte X25519 public key followed by the 1,088-byte ML-KEM-768 ciphertext~\cite{fips203}. The other two pairs follow the same pattern with their respective NIST-specified key and ciphertext lengths.
\newline\noindent\textbf{TCP socket plugin}: As per RFC~9329~\cite{rfc9329}, each IKE datagram is wrapped in a 4-byte big-endian length prefix and sent over a TCP stream on port~4500. 
The plugin intercepts \texttt{charon}'s send/receive path at the socket abstraction layer. 
The initiator side establishes the TCP connection before \textsc{ike\_sa\_init} begins, and the responder side binds a listening socket and accepts the incoming connection. 
The plugin also maintains the keepalive timers
and supports connection pooling so that a rekeyed IKE SA can reuse the existing TCP session without a new three-way handshake. Note that the TCP connection setup now adds a three-way handshake (connection setup) overhead to the total handshake time, regardless of the payload size or loss rate.
\newline\noindent\textbf{Separate\_Transports negotiation}:
When the initiator begins \textsc{ike\_sa\_init} over TCP, the plugin automatically appends a \textsc{separate\_transports} notify payload (type 16460) as defined in the draft~\cite{reddydraft}. 
The responder plugin checks for this notification and echoes it back in the response. 
After \textsc{ike\_auth} completes with the echo present, both peers
install the Child SA with the IPsec policy set to UDP for ESP~\cite{rfc4303}, while all subsequent IKEv2~\cite{rfc7296} control messages \textit{e.g.}, rekeying, liveness checks INFORMATIONAL exchanges continue over the established TCP connection. 
This split is transparent to the application, where ESP packets use the standard UDP port~4500. 
In the absence of an echo of textsc{separate\_transports}, the initiator falls back to a single TCP transport making both IKE and ESP~\cite{rfc4303} use the same TCP as defined in RFC~9329~\cite{rfc9329}.

\noindent\textbf{Universal Combiner:}
After the IKE payload exchange, each peer independently derives two raw shared secrets as discussed in \S~\ref{sec:background:draft}. 
The HKDF-based Universal Combiner~\cite{rfc5869, hybridkems} is then called with these two secrets, the ML-KEM ciphertext $\mathit{CT}_{PQ}$, both public keys, and the protocol-specific label \texttt{"IKEv2-HYBRID-KE-v1"} encoded as a length-prefixed byte sequence. 
The output replaces the $g^{ir}$ term in the standard SKEYSEED computation~\cite{rfc7296}.
Derivation of all other keying material (\textit{SK\_d}, \textit{SK\_ai/ar}, \textit{SK\_ei/er}) remains the same (as specified in RFC~7296~\cite{rfc7296}).

\input{figures/CompositeExecution}

\noindent\textbf{Certificates and authentication}:
We generate four certificate chains to cover the authentication matrix:
RSA-2048, ML-DSA-44 (baseline), ML-DSA-65, and ML-DSA-87~\cite{fips204},
all signed by a single CA. strongSwan~\cite{strongSwan} selects the
authentication algorithm from the peer's certificate at \textsc{ike\_auth} time, and no configuration change is needed between runs.

The Algorithm~\ref{fig:composite_algo} describes the step-by-step execution detailing the sequence of interactions between the Initiator (client) and Responder (Server) components. 
This algorithm presents the execution flow of our implementation, detailing the fixed-offset payload concatenation (\texttt{composite-KE}), the RFC~9329 message framing (\texttt{tcp-socket}), the datagram steering (\texttt{separate\_transports}), and the Universal Combiner key derivation.

%% file: figures/CompositeExecution.tex
\begin{figure*}[t]
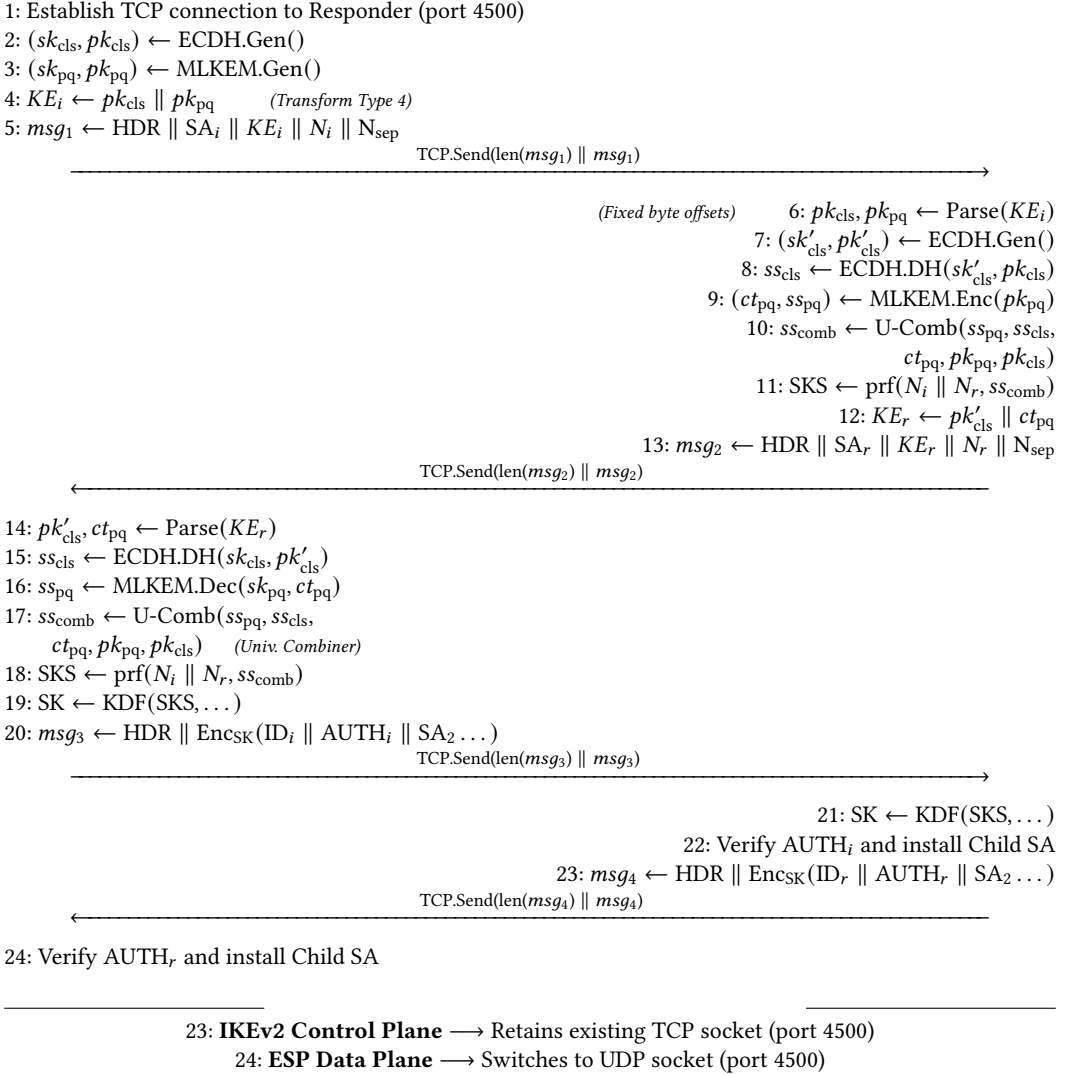

\hrule height 0.8pt
\vspace{4pt}
\textbf{Algorithm} Detailed Protocol Execution: Composite Key Exchange over TCP
\vspace{2pt}
\hrule height 0.4pt
\vspace{6pt}
\small
\begin{tabular*}{\linewidth}{@{}l@{\extracolsep{\fill}}r@{}}
\framebox{\textbf{Initiator}} & \framebox{\textbf{Responder}} \\
\noalign{\vspace{4pt}}
1: Establish TCP connection to Responder (port 4500) & \\
2: $(sk_{\text{cls}}, pk_{\text{cls}}) \leftarrow \text{ECDH.Gen}()$ & \\
3: $(sk_{\text{pq}}, pk_{\text{pq}}) \leftarrow \text{MLKEM.Gen}()$ & \\
4: $KE_i \leftarrow pk_{\text{cls}} \parallel pk_{\text{pq}}$ \hspace{2em}\textit{\scriptsize (Transform Type 4)} & \\
5: $msg_1 \leftarrow \text{HDR} \parallel \text{SA}_i \parallel KE_i \parallel N_i \parallel \text{N}_{\text{sep}}$ & \\
\multicolumn{2}{@{}c@{}}{$\xrightarrow{\makebox[12cm]{\scriptsize TCP.Send(\text{len}($msg_1$) $\parallel msg_1$)}}$} \\
\noalign{\vspace{4pt}}
& \textit{\scriptsize (Fixed byte offsets)}\hspace{2em} 6: $pk_{\text{cls}}, pk_{\text{pq}} \leftarrow \text{Parse}(KE_i)$ \\
& 7: $(sk'_{\text{cls}}, pk'_{\text{cls}}) \leftarrow \text{ECDH.Gen}()$ \\
& 8: $ss_{\text{cls}} \leftarrow \text{ECDH.DH}(sk'_{\text{cls}}, pk_{\text{cls}})$ \\
& 9: $(ct_{\text{pq}}, ss_{\text{pq}}) \leftarrow \text{MLKEM.Enc}(pk_{\text{pq}})$ \\
& 10: $ss_{\text{comb}} \leftarrow \text{U-Comb}(ss_{\text{pq}}, ss_{\text{cls}},$ \\
& \qquad $ct_{\text{pq}}, pk_{\text{pq}}, pk_{\text{cls}})$ \\
& 11: $\text{SKS} \leftarrow \text{prf}(N_i \parallel N_r, ss_{\text{comb}})$ \\
& 12: $KE_r \leftarrow pk'_{\text{cls}} \parallel ct_{\text{pq}}$ \\
& 13: $msg_2 \leftarrow \text{HDR} \parallel \text{SA}_r \parallel KE_r \parallel N_r \parallel \text{N}_{\text{sep}}$ \\
\multicolumn{2}{@{}c@{}}{$\xleftarrow{\makebox[12cm]{\scriptsize TCP.Send(\text{len}($msg_2$) $\parallel msg_2$)}}$} \\
\noalign{\vspace{4pt}}
14: $pk'_{\text{cls}}, ct_{\text{pq}} \leftarrow \text{Parse}(KE_r)$ & \\
15: $ss_{\text{cls}} \leftarrow \text{ECDH.DH}(sk_{\text{cls}}, pk'_{\text{cls}})$ & \\
16: $ss_{\text{pq}} \leftarrow \text{MLKEM.Dec}(sk_{\text{pq}}, ct_{\text{pq}})$ & \\
17: $ss_{\text{comb}} \leftarrow \text{U-Comb}(ss_{\text{pq}}, ss_{\text{cls}},$ & \\
\qquad $ct_{\text{pq}}, pk_{\text{pq}}, pk_{\text{cls}})$ \hspace{1em}\textit{\scriptsize (Univ. Combiner)} & \\
18: $\text{SKS} \leftarrow \text{prf}(N_i \parallel N_r, ss_{\text{comb}})$ & \\
19: $\text{SK} \leftarrow \text{KDF}(\text{SKS}, \dots)$ & \\
20: $msg_3 \leftarrow \text{HDR} \parallel \text{Enc}_{\text{SK}}(\text{ID}_i \parallel \text{AUTH}_i \parallel \text{SA}_2 \dots)$ & \\
\multicolumn{2}{@{}c@{}}{$\xrightarrow{\makebox[12cm]{\scriptsize TCP.Send(\text{len}($msg_3$) $\parallel msg_3$)}}$} \\
\noalign{\vspace{4pt}}
& 21: $\text{SK} \leftarrow \text{KDF}(\text{SKS}, \dots)$ \\
& 22: Verify $\text{AUTH}_i$ and install Child SA \\
& 23: $msg_4 \leftarrow \text{HDR} \parallel \text{Enc}_{\text{SK}}(\text{ID}_r \parallel \text{AUTH}_r \parallel \text{SA}_2 \dots)$ \\
\multicolumn{2}{@{}c@{}}{$\xleftarrow{\makebox[12cm]{\scriptsize TCP.Send(\text{len}($msg_4$) $\parallel msg_4$)}}$} \\
\noalign{\vspace{4pt}}
24: Verify $\text{AUTH}_r$ and install Child SA & \\
\noalign{\vspace{6pt}}
\hrulefill & \hrulefill \\
\multicolumn{2}{@{}c@{}}{23: \textbf{IKEv2 Control Plane} $\longrightarrow$ Retains existing TCP socket (port 4500)} \\
\multicolumn{2}{@{}c@{}}{24: \textbf{ESP Data Plane} $\longrightarrow$ Switches to UDP socket (port 4500)} \\
\end{tabular*}
\vspace{4pt}
\hrule height 0.8pt
\vspace{2ex}
\caption{Detailed cryptographic execution of the composite key exchange over TCP, including payload concatenation and Universal Combiner key derivation. $\text{N}_{\text{sep}}$ denotes the \textsc{separate\_transports} payload.}
\label{fig:composite_algo}
\end{figure*}

%% file: sections/04_setup.tex
\section{Evaluation}
\label{sec:eval}
In this section, we discuss the hardware and software configuration for our setup to run and evaluate IPSec. We provide details on the crypto primitives used, the network conditions under which they are run, and a list of metrics for comparing their performance. 

\subsection{Experimental Setup: Hardware and Software}
We set up two Docker containers on Ubuntu~24.04 with Linux kernel~6.17.0-29, running on a Intel Core i7-14700K host machine with 32\,GB DDR4 memory. 
One container serves as an initiator and the other as a responder. 
To ensure consistent performance, the CPU frequency was fixed at 3.4,GHz using the \texttt{performance} governor mode. Further, we disabled Intel Turbo Boost, and package C-states were restricted to C1 at best.
These settings ensure that Read Time-Stamp Counter ~(RDTSC)~\cite{rdtsc_paoloni} cycle counts accurately reflect the elapsed wall-clock time while enabling reproducible Intel Running Average Power Limit (RAPL)~\cite{rapl} readings across multiple measurements.
The software stack consists of strongSwan~6.0~\cite{strongSwan} (\texttt{ml-dsa}
dev. branch), liboqs~0.8.0, oqs-provider~0.5.0 with OpenSSL~3.x, and Docker~24.0.7. All containers were built from the same base image to guarantee identical versions of cryptographic library across the Initiator and Responder.

\subsection{Testbed Architecture}
Figure~\ref{fig:testbed} shows the testbed used to establish IPsec connections. The main components of the testbed and relevant configurations used to establish the IPsec connections are described below.
\newline\noindent\textbf{Setup on Host Node {\small\normalfont(\circnum{1})}:}
Before starting the containers, the host script fixes the CPU frequency at the P-core base clock, disables Intel Turbo Boost, restricts package C-states to at most C1, and sets the CPU frequency governor to \texttt{performance}.
After applying these settings, the script reads the current CPU frequency from \texttt{sysfs} and aborts if it deviates from the target value. The same verification is repeated after the measurement window to detect any thermal throttling or unintended frequency scaling that may have occurred during the experiment.
\newline\noindent\textbf{Certificate {\small\normalfont(\circnum{2})}:}
The Responder generates all three X.509 certificates at container startup: a self-signed CA, a Responder leaf certificate, and an Initiator leaf certificate. All three are written to a shared Docker volume. The Initiator polls the volume and copies its credentials before starting the IKEv2 daemon. Certificate generation time is excluded from all handshake measurements because both daemons are already loaded and configured by the time the first \textsc{ike\_sa\_init} is sent.
\newline\noindent\textbf{Connection setup {\small\normalfont(\circnum{3} \circnum{5} \circnum{7} \circnum{9}):}} 
The Initiator orchestrates the connection loop. 
Each measurement window lasts 60 seconds, during which handshake attempts are performed continuously while \texttt{tcpdump} records all packets exchanged between the initiator and responder.
After the measurement window ends, \texttt{tcpdump} is terminated, and \texttt{tshark} analyzes the resulting packet capture to determine the actual \textsc{ike\_sa\_init} payload sizes and total bytes exchanged per handshake.
\newline\noindent\textbf{Measurements {\small\normalfont(\circnum{14} \circnum{15} \circnum{16}):}}
For each handshake, Intel RAPL energy is sampled at the Package domain immediately before and after each IKE initiation. RDTSC hardware timestamps bracket the same interval. Counter wrap-around is handled per-domain to ensure correctness across long runs. After each successful handshake, the IKE SA is torn down and the SA table is periodically flushed to prevent accumulated state from affecting subsequent measurements under high loss.
\newline\noindent\textbf{Plugin stack {\small\normalfont(\circnum{6} \circnum{10})}:}
The \texttt{charon} daemon on each side loads the plugins required for the current run. For UDP runs, \texttt{socket-default} is active and \texttt{tcp-socket} is disabled. For TCP runs, the roles are reversed and \texttt{separate\_transports} is loaded so that IKE travels over TCP while ESP continues over UDP. Composite algorithm runs additionally load the \texttt{composite-KE} plugin. The transport assignment for each run is set explicitly at experiment launch and does not change during the run.

\input{figures/fig2_testbed}

\noindent\textbf{Software Bridge {\small\normalfont(\circnum{11})} and Network impairment {\small\normalfont(\circnum{12})}:}
Both containers share a dedicated Docker bridge network (\texttt{ipsec\_net}, \texttt{172.21.0.0/24}, MTU~1500\,B by default), which carries all IKEv2 control-plane and ESP data-plane traffic. The bridge interface also serves as the passive tap point for \texttt{tcpdump}: packet capture runs on this shared interface, so both the Initiator's egress and the Responder's egress are visible in a single capture file without requiring promiscuous mode on either container's veth pair. Network impairment is applied unidirectionally: the \texttt{tc/netem} module is attached to the Initiator's egress interface inside the container, injecting configurable delay and packet-loss, while the Responder's return path remains unimpaired. This design isolates the effect of forward-path loss on IKEv2 retransmission behavior without introducing correlated loss on both directions simultaneously.



\subsection{Measurement Execution Flow}
\label{sec:flow}
Each experiment proceeds as follows.
The host performance script (\circnum{1}) locks the CPU before any container is launched.
The Responder container starts first: it generates X.509 certificates into a shared volume, loads the \texttt{charon} daemon with the plugin set for the current algorithm family, and begins listening on UDP/4500 (or TCP/4500 for composite runs).
The Initiator container starts next, polls the shared volume until credentials are present, and then enters a warm-up phase of 15 consecutive IKEv2 handshakes whose results are discarded.
Once warm-up completes, \texttt{tcpdump} is started on the \texttt{ipsec\_net} bridge interface (\circnum{11}) to capture all subsequent traffic, and the 60-second measurement window begins.
For each handshake in the window, the Initiator sends an \textsc{ike\_sa\_init} request over the bridge; for hybrid algorithms this is followed by an \textsc{ike\_intermediate} exchange carrying the PQC key share, while composite runs bundle both classical and PQ key material into the initial \textsc{ike\_sa\_init} payload
In both cases, \textsc{ike\_auth} finalises mutual authentication via ML-DSA and establishes the Child SA.
RAPL energy (\circnum{14}) and RDTSC cycle counts (\circnum{15}) are sampled immediately before and after each \textsc{ike\_sa\_init} transmission inside the patched strongSwan daemon.
After the 60-second window closes, \texttt{tcpdump} is stopped, \texttt{tshark} parses the capture (\circnum{16}) to extract per-handshake wire-byte counts, and results are written to a timestamped CSV for post-processing.
\subsection{Retransmission configuration}
Retransmission time significantly affects the performance of IPSec tunnels.  strongSwan's~\cite{strongSwan} retransmit parameters are configured with a base timeout of 1.0,s, a multiplier of 1.2, and a maximum of three attempts, resulting in the backoff sequence 1.0,s, 1.2,s, and 1.44,s. 
These values comply with the retransmission requirements in RFC~7296~\cite{rfc7296}, which recommends an initial retransmission interval of at least one second and bounds the maximum wait time for a failed handshake to 3.64,s.
The 1.0\,s base retransmission interval exposes loss-induced latency effects while avoiding extreme outliers that could distort the scale of the Composite-versus-Hybrid comparison.
All retransmit parameters are configured identically for both TCP and UDP, ensuring that any observed latency differences can be attributed to transport-layer behaviour rather than the timer configuration.

%% file: figures/fig2_testbed.tex
\begin{figure}[t]
\centering
\IfFileExists{figures/pdf/Figure3-Testbed.pdf}{%
  \includegraphics[width=\linewidth]{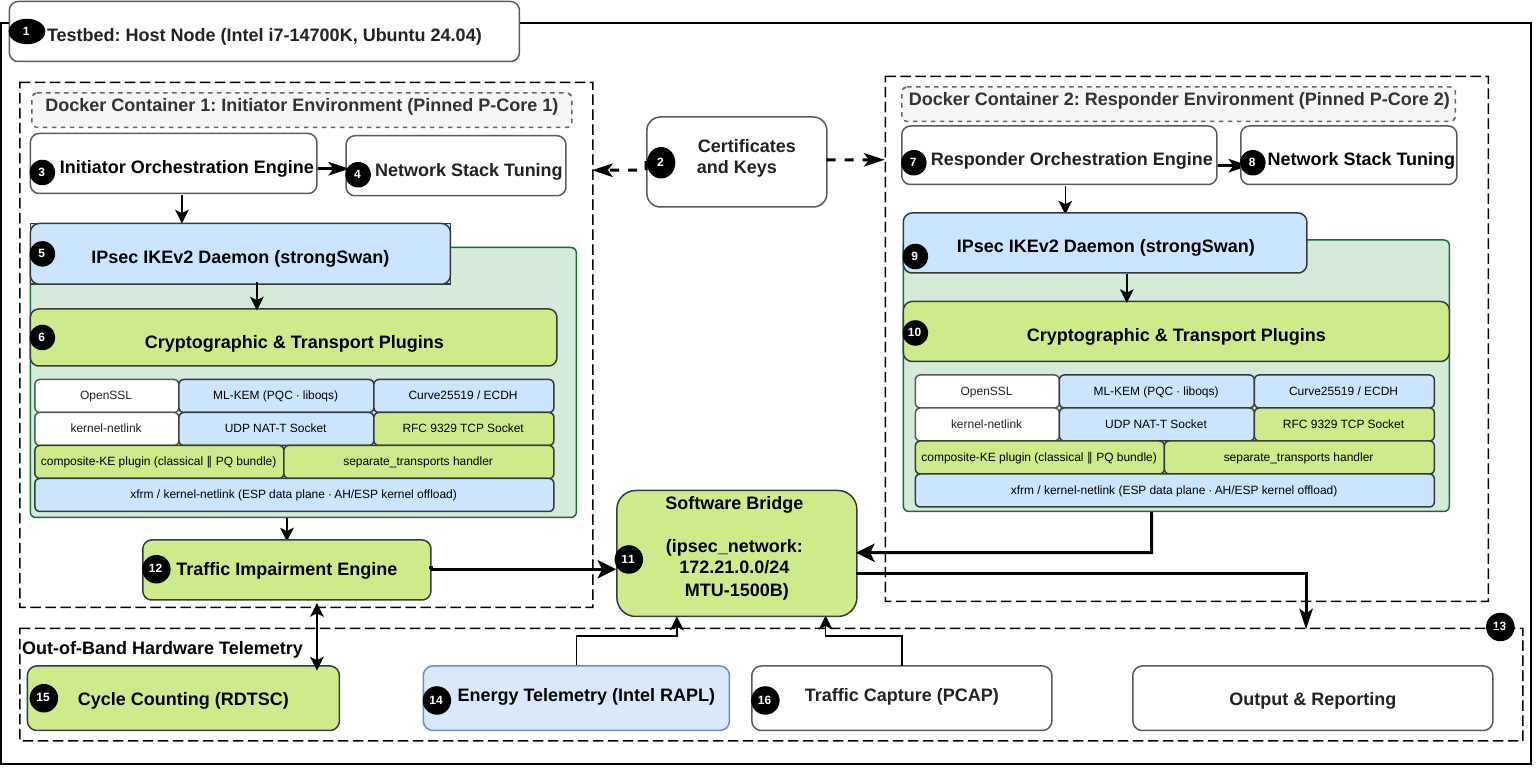}%
}{%
  \fbox{\parbox{0.92\linewidth}{\centering Put \texttt{figures/fig2\_testbed.png} here.}}%
}
\caption{Testbed architecture: Two Docker containers pinned to dedicated CPU cores of host node and connected through a Docker bridge with configurable link delay and loss parameters. The components are color-coded to denote the implementation effort: white boxes represent unmodified standard tools (Used As-Is), blue boxes indicate patched open-source components (Modified/Extended), and green boxes indicate the newly developed components added as part of this work (Built from Scratch).}

\label{fig:testbed}
\end{figure}

%% file: sections/05_algorithms.tex
\subsection{Crypto primitives Under Test}
\label{sec:CryptoPrimitives}

We evaluate the performance of six key-exchange configurations for encapsulation. Three of them are precisely the Composite transform pairs enumerated in draft~\cite{reddydraft}. Their three Hybrid counterparts use the exact same cryptographic combinations pairing the classical Curve25519, ECP-256, and ECP-384 groups with post-quantum ML-KEM-768 and ML-KEM-1024. This covers two distinct security levels (NIST Level~1/2 and Level~5). The evaluation is deliberately narrow so that transport behavior is in focus, not crypto primitives diversity, and remains the variable of interest. All six key- exchange methods are evaluated against post quantum digital signature algorithms ML-DSA-44~\cite{fips204} (FIPS~204 Level~2), ML-DSA-65 (Level~3) and ML-DSA-87 (Level~5)
%
\newline\noindent\textbf{Composite (C):}
Composite algorithms concatenate the classical and post-quantum public keys into a single \textsc{ke} payload transmitted in \textsc{ike\_sa\_init}~\cite{reddydraft}. The three pairs in scope (Curve25519$\Vert$ML-KEM-768, ECP-256$\Vert$ML-KEM-768, and ECP-384$\Vert$ML-KEM-1024) produce
\textsc{ike\_sa\_init} payloads of 1,434, 1,466, and 1,882 bytes, respectively. The (ECP-384$\Vert$ML-KEM-1024) exceeds the 1,500-byte Ethernet MTU, forcing fragmentation over UDP. 
%
\newline\noindent\textbf{Hybrid (H):}
In the hybrid approach \textsc{ike\_sa\_init} carries only the classical key share, deferring the post-quantum key to a separate \textsc{ike\_intermediate} (ADDKE) exchange~\cite{rfc9370}. 
Correspondingly, the first message is small (256, 288, and 330 bytes for Curve25519$+$ML-KEM-768, ECP-256$+$ML-KEM-768, and ECP-384$+$ML-KEM-1024, respectively) and fits within a single UDP datagram. 
The deferred ML-KEM public key is then transmitted in an \textsc{ike\_intermediate} request ($\approx$1{,}316\,B), with the ciphertext returned in the response ($\approx$1{,}180\,B). 
Both these messages may engender fragmentation~\cite{rfc7383}.
The hybrid mechanism avoids using TCP at the cost of one additional round trip. 
%

Throughout the paper, Curve25519$+$ML-KEM-768\,(H/UDP) serves as the primary baseline. 
This configuration represents the most conservative compliant deployment, as this does not require TCP, and uses the smallest classical key-exchange group considered in our evaluation. 


\subsection{Network Scenario Matrix}

Each composite and hybrid key-sharing method is tested across various network conditions by varying the packet loss and packet delays using linux native tc/netm command. packet loss is varied from 0\% to 10\% and network delays from $0$ ms to $10$ms. The packet loss range spans what enterprise LANs see at the low end and what degraded satellite or mobile paths see at the high end~\cite{huang2012lte,michel2022starlink}. The packet delay range isolate the round-trip-count effect without the compounding influence of retransmission timers. 


\subsection{Performance Metrics}
\label{sec:setup:metrics}
For every handshake, we record the following five parameters:
\begin{itemize}
    \item \textbf{Handshake Latency:} At the intiator, we measure the Wall-clock time starting from the \texttt{swanctl --initiate} till the reception of SA confirmation. The window encompasses all IKE message exchanges, retransmissions, key derivation, and Child SA installation. Under packet-loss conditions, this metric most directly reflects the perceived experience of an end user or a VPN gateway.
    \item \textbf{Throughput:}
    The number of successfully completed IKE SAs per second during the 60-second measurement window. This metric captures how packet loss affects the performance of VPN gateway beyond the per-handshake latency.  
    \item \textbf{CPU Cycles:} The number of processor cycles consumed per handshake, measured at the initiator using \texttt{RDTSC} with a preceding \texttt{CPUID} serialization fence to ensure accurate cycle-count measurements.
    Core pinning enables more precise measurements of cryptographic and protocol processing overhead by isolating the workload from operating-system scheduler-induced variability.
    \item \textbf{Energy:} We measure the CPU package-domain energy consumed per handshake. This is derived from Intel RAPL~\cite{rapl} measurements collected on the host machine. Collecting energy measurements outside the containers avoids virtualisation-induced measurement artefacts. The per-handshake value is computed as the cumulative energy difference over the measurement interval divided by the number of successfully completed handshakes.
    \item \textbf{IKE Wire Bytes:} We measure the total bidirectional IKE-layer bytes exchanged per successful handshake. We derive this from \texttt{tcpdump} captures on the bridge interface that both containers share. This metric directly captures the overhead on wire imposed by larger post-quantum payloads.
\end{itemize}

%% file: sections/06_results.tex

\section{Results}
\label{sec:results}

In this section we report the performance IKEv2 handshakes with various cypto-primitives discussed in Section~\ref{sec:CryptoPrimitives} under various network conditions. All connections are established using the same physical machine and software stack. For each combination of crypto-primitives and network conditions, every new connection is established as an independent connection.
Before any collection of data, 15 warm-up handshakes are performed. The warm-up handshakes are increased to 30 when a lossy netem profile is active to account for inflated variance during the early transient phase. 
Warm-up connections are terminated and discarded without writing any output. After warm-up, packet capture starts on the bridge interface to prevent warm-up traffic from being mixed into wire-byte counts. The measurement window then runs for $60$ seconds, during which continuous handshake attempts are made serially one after another. After the 60-second window closes, \texttt{tcpdump} is stopped and \texttt{tshark} parses the capture file to extract the real \textsc{ike\_sa\_init} payload sizes and total per-handshake byte counts.

Table~\ref{tab:compvshyb2} and Table~\ref{tab:energy_results} records the main result of this section. 
they report handshake latency and energy consumption for all three key-exchange variants
(Composite/TCP, Hybrid/TCP, Hybrid/UDP) across various network scenarios
in conjunction with two Digital Signature Algorithms  -- ML-DSA-44 and ML-DSA-65. Detailed comparison with  other RSA-2048, ML-DSA-44,
ML-DSA-65, and ML-DSA-87 appears in Section~\ref{sec:results:auth}. Supporting table on the throughput and CPU cycles are given in the Appendix.

Two effects dominate our observation across varied network conditions -- the key-exchange variants transport mechanism and the Digital Signature Algorithm (DSA). Hybrid (H/UDP)
degrades exponentially with packet loss while both TCP variants (H/TCP, C/TCP),  degrade
gradually, because UDP exposes the IKEv2 retransmission timer directly to
fragment loss whereas TCP absorbs the same loss below the IKE state machine.
Switching from ML-DSA-44 to ML-DSA-65 adds an approximately constant increment in latency, CPU cycles, and energy
under different network conditions. The effect of transport protocol choice and the digital signature algorithm are independent
and additive. In the following we treat them separately. For exploration of the complete benchmarking result across all tested algorithms and network scenarios, an web dashboard is available at~\cite{anon_dashboard}.

\input{sections/06_table2_compvshyb}
\input{sections/06_energy_table}
\input{sections/06a_ideal}
\input{sections/06b_under_loss}
\input{sections/06c_rto_analysis}
\input{sections/06d_under_delay}
\input{sections/06e_combined}
\input{sections/06g_energy}
\input{sections/06h_auth_comparison}
\input{sections/07_wire_bytes}


%% file: sections/06_table2_compvshyb.tex

\newcommand{\barcell}[3]{%
  \pgfmathparse{(#1)/(#3)*46}\let\bwpt\pgfmathresult%
  \begin{tikzpicture}[x=1pt,y=1pt,baseline=2pt]
    \useasboundingbox (0,-1) rectangle (46,8);
    \draw[cyan!60!black, line width=0.25pt] (0,0) rectangle (46,7);
    \fill[cyan!30!white] (0.2,0.2) rectangle (\bwpt,6.8);
    \node[anchor=east, inner sep=1pt, font=\scriptsize] at (45.5,3.5) {#1($\pm$#2)};
  \end{tikzpicture}%
}

\newcommand{\nabar}{%
  \begin{tikzpicture}[x=1pt,y=1pt,baseline=2pt]
    \useasboundingbox (0,-1) rectangle (46,8);
    \draw[cyan!60!black, line width=0.25pt] (0,0) rectangle (46,7);
    \node[anchor=east, inner sep=1pt, font=\scriptsize] at (45.5,3.5) {---};
  \end{tikzpicture}%
}

\newcommand{\maxCrvC}{67.2}
\newcommand{\maxCrvHT}{91.4}
\newcommand{\maxCrvHU}{918.0}

\newcommand{\maxEcpC}{60.1}
\newcommand{\maxEcpHT}{93.6}
\newcommand{\maxEcpHU}{890.8}

\newcommand{\maxEtcC}{56.7}
\newcommand{\maxEtcHT}{97.3}
\newcommand{\maxEtcHU}{923.0}

\newcommand{\CcCrv}[2]{\barcell{#1}{#2}{\maxCrvC}}
\newcommand{\HTCrv}[2]{\barcell{#1}{#2}{\maxCrvHT}}
\newcommand{\HUCrv}[2]{\barcell{#1}{#2}{\maxCrvHU}}

\newcommand{\CcEcp}[2]{\barcell{#1}{#2}{\maxEcpC}}
\newcommand{\HTEcp}[2]{\barcell{#1}{#2}{\maxEcpHT}}
\newcommand{\HUEcp}[2]{\barcell{#1}{#2}{\maxEcpHU}}

\newcommand{\CcEtc}[2]{\barcell{#1}{#2}{\maxEtcC}}
\newcommand{\HTEtc}[2]{\barcell{#1}{#2}{\maxEtcHT}}
\newcommand{\HUEtc}[2]{\barcell{#1}{#2}{\maxEtcHU}}

\newcommand{\scendash}{\noalign{\vskip1pt}\cdashline{1-11}[0.5pt/1.5pt]\noalign{\vskip1pt}}

\begin{table*}[t]
\centering
\caption{Handshake latency (ms) with standard deviations (inside brackets) across transport modes and algorithm pairs. Bars represent latency normalized to the column maximum.}
\label{tab:compvshyb2}
\setlength{\tabcolsep}{3pt}
\renewcommand{\arraystretch}{1.1}
\scriptsize
\resizebox{\textwidth}{!}{%
\begin{tabular}{@{}ll
  l@{\hspace{2pt}}l@{\hspace{2pt}}l|
  l@{\hspace{2pt}}l@{\hspace{2pt}}l|
  l@{\hspace{2pt}}l@{\hspace{2pt}}l@{}}
\toprule
& & \multicolumn{3}{c|}{\textbf{(Curve25519, ML-KEM-768)}}
  & \multicolumn{3}{c|}{\textbf{(ECP-256, ML-KEM-768)}}
  & \multicolumn{3}{c}{\textbf{(ECP-384, ML-KEM-1024)}} \\
\textbf{Scenario} & \textbf{DSA}
  & \textbf{Comp(TCP)} & \textbf{Hyb(TCP)} & \textbf{Hyb(UDP)}
  & \textbf{Comp(TCP)} & \textbf{Hyb(TCP)} & \textbf{Hyb(UDP)}
  & \textbf{Comp(TCP)} & \textbf{Hyb(TCP)} & \textbf{Hyb(UDP)} \\
\midrule
\multirow{2}{*}{0\%}
  & ML-DSA-44  & \CcCrv{11.6}{0.1} & \HTCrv{12.0}{0.1} & \HUCrv{11.7}{0.1} & \CcEcp{11.8}{0.1} & \HTEcp{12.2}{0.1} & \HUEcp{11.9}{0.1} & \CcEtc{15.4}{0.1} & \HTEtc{16.0}{0.1} & \HUEtc{15.8}{0.1} \\
  & ML-DSA-65 & \CcCrv{14.5}{0.1} & \HTCrv{14.9}{0.1} & \HUCrv{14.5}{0.1} & \CcEcp{14.8}{0.1} & \HTEcp{15.1}{0.1} & \HUEcp{14.8}{0.1} & \CcEtc{18.4}{0.1} & \HTEtc{19.0}{0.1} & \HUEtc{18.6}{0.1} \\
\scendash
\multirow{2}{*}{0.25\%}
  & ML-DSA-44  & \CcCrv{12.8}{0.2} & \HTCrv{13.8}{0.2} & \HUCrv{22.5}{1.5} & \CcEcp{12.3}{0.2} & \HTEcp{13.7}{0.2} & \HUEcp{22.4}{1.5} & \CcEtc{16.1}{0.2} & \HTEtc{17.3}{0.2} & \HUEtc{28.2}{1.8} \\
  & ML-DSA-65 & \CcCrv{15.3}{0.2} & \HTCrv{16.5}{0.2} & \HUCrv{36.4}{2.1} & \CcEcp{15.4}{0.2} & \HTEcp{17.1}{0.2} & \HUEcp{35.6}{2.0} & \CcEtc{19.0}{0.2} & \HTEtc{20.3}{0.2} & \HUEtc{39.1}{2.2} \\
\scendash
\multirow{2}{*}{0.5\%}
  & ML-DSA-44  & \CcCrv{13.8}{0.3} & \HTCrv{14.9}{0.3} & \HUCrv{32.9}{1.8} & \CcEcp{12.9}{0.2} & \HTEcp{15.0}{0.3} & \HUEcp{38.2}{2.1} & \CcEtc{16.7}{0.3} & \HTEtc{19.1}{0.4} & \HUEtc{38.1}{2.0} \\
  & ML-DSA-65 & \CcCrv{16.1}{0.3} & \HTCrv{18.6}{0.4} & \HUCrv{58.2}{3.2} & \CcEcp{15.7}{0.3} & \HTEcp{18.0}{0.4} & \HUEcp{64.7}{3.5} & \CcEtc{19.5}{0.4} & \HTEtc{21.7}{0.4} & \HUEtc{69.9}{3.8} \\
\scendash
\multirow{2}{*}{1\%}
  & ML-DSA-44  & \CcCrv{15.7}{0.4} & \HTCrv{19.0}{0.5} & \HUCrv{58.8}{3.1} & \CcEcp{13.9}{0.3} & \HTEcp{18.8}{0.5} & \HUEcp{60.9}{3.2} & \CcEtc{17.8}{0.4} & \HTEtc{22.5}{0.6} & \HUEtc{69.3}{3.6} \\
  & ML-DSA-65 & \CcCrv{18.2}{0.4} & \HTCrv{22.1}{0.6} & \HUCrv{118.4}{5.8} & \CcEcp{17.2}{0.4} & \HTEcp{21.7}{0.6} & \HUEcp{102.2}{5.2} & \CcEtc{20.4}{0.5} & \HTEtc{26.1}{0.7} & \HUEtc{111.5}{5.5} \\
\scendash
\multirow{2}{*}{2\%}
  & ML-DSA-44  & \CcCrv{20.7}{0.6} & \HTCrv{25.4}{0.8} & \HUCrv{106.2}{4.8} & \CcEcp{16.5}{0.5} & \HTEcp{25.1}{0.8} & \HUEcp{117.2}{5.2} & \CcEtc{20.1}{0.6} & \HTEtc{29.1}{0.9} & \HUEtc{114.6}{5.0} \\
  & ML-DSA-65 & \CcCrv{23.8}{0.7} & \HTCrv{28.4}{0.9} & \HUCrv{218.8}{9.5} & \CcEcp{20.0}{0.6} & \HTEcp{28.3}{0.9} & \HUEcp{191.4}{8.5} & \CcEtc{23.2}{0.7} & \HTEtc{33.1}{1.0} & \HUEtc{234.9}{9.8} \\
\scendash
\multirow{2}{*}{5\%}
  & ML-DSA-44  & \CcCrv{33.6}{1.1} & \HTCrv{44.8}{1.5} & \HUCrv{255.6}{11.2} & \CcEcp{25.3}{0.8} & \HTEcp{45.5}{1.6} & \HUEcp{302.6}{13.1} & \CcEtc{27.5}{0.9} & \HTEtc{45.6}{1.5} & \HUEtc{261.6}{11.4} \\
  & ML-DSA-65 & \CcCrv{36.7}{1.2} & \HTCrv{51.4}{1.7} & \HUCrv{474.4}{19.5} & \CcEcp{30.3}{1.0} & \HTEcp{52.7}{1.8} & \HUEcp{471.9}{19.8} & \CcEtc{32.7}{1.1} & \HTEtc{52.7}{1.7} & \HUEtc{459.0}{18.2} \\
\scendash
\multirow{2}{*}{10\%}
  & ML-DSA-44  & \CcCrv{58.0}{2.0} & \HTCrv{83.0}{3.1} & \HUCrv{582.1}{24.3} & \CcEcp{43.3}{1.5} & \HTEcp{82.1}{2.9} & \HUEcp{548.4}{23.1} & \CcEtc{51.6}{1.8} & \HTEtc{86.8}{3.2} & \HUEtc{575.8}{24.0} \\
  & ML-DSA-65 & \CcCrv{67.2}{2.4} & \HTCrv{91.4}{3.5} & \HUCrv{918.0}{38.2} & \CcEcp{60.1}{2.1} & \HTEcp{93.6}{3.4} & \HUEcp{890.8}{37.5} & \CcEtc{56.7}{2.0} & \HTEtc{97.3}{3.6} & \HUEtc{923.0}{38.8} \\
\midrule
\multirow{2}{*}{1\,ms/0\%}
  & ML-DSA-44  & \CcCrv{13.5}{0.1} & \HTCrv{14.9}{0.1} & \HUCrv{14.7}{0.1} & \CcEcp{13.8}{0.1} & \HTEcp{15.3}{0.1} & \HUEcp{15.0}{0.1} & \CcEtc{17.4}{0.1} & \HTEtc{19.1}{0.1} & \HUEtc{18.8}{0.1} \\
  & ML-DSA-65 & \CcCrv{16.6}{0.1} & \HTCrv{17.9}{0.1} & \HUCrv{17.7}{0.1} & \CcEcp{16.9}{0.1} & \HTEcp{18.2}{0.1} & \HUEcp{18.0}{0.1} & \CcEtc{20.6}{0.1} & \HTEtc{22.1}{0.1} & \HUEtc{21.8}{0.1} \\
\scendash
\multirow{2}{*}{5\,ms/0\%}
  & ML-DSA-44  & \CcCrv{21.3}{0.2} & \HTCrv{26.8}{0.2} & \HUCrv{26.8}{0.2} & \CcEcp{21.6}{0.2} & \HTEcp{27.1}{0.2} & \HUEcp{27.0}{0.2} & \CcEtc{25.1}{0.2} & \HTEtc{30.9}{0.2} & \HUEtc{30.8}{0.2} \\
  & ML-DSA-65 & \CcCrv{24.0}{0.2} & \HTCrv{29.5}{0.2} & \HUCrv{29.4}{0.2} & \CcEcp{24.4}{0.2} & \HTEcp{29.7}{0.2} & \HUEcp{29.7}{0.2} & \CcEtc{28.1}{0.2} & \HTEtc{33.5}{0.2} & \HUEtc{33.3}{0.2} \\
\scendash
\multirow{2}{*}{10\,ms/0\%}
  & ML-DSA-44  & \CcCrv{31.4}{0.3} & \HTCrv{41.8}{0.4} & \HUCrv{41.7}{0.4} & \CcEcp{31.6}{0.3} & \HTEcp{42.1}{0.4} & \HUEcp{42.1}{0.4} & \CcEtc{35.1}{0.3} & \HTEtc{45.9}{0.4} & \HUEtc{45.7}{0.4} \\
  & ML-DSA-65 & \CcCrv{34.1}{0.3} & \HTCrv{44.5}{0.4} & \HUCrv{44.4}{0.4} & \CcEcp{34.4}{0.3} & \HTEcp{44.7}{0.4} & \HUEcp{44.7}{0.4} & \CcEtc{38.1}{0.4} & \HTEtc{48.7}{0.5} & \HUEtc{48.6}{0.5} \\
\midrule
2\,ms/2\%
  & ML-DSA-65 & \CcCrv{28.6}{0.9} & \HTCrv{34.6}{1.2} & \HUCrv{193.3}{8.2} & \CcEcp{23.4}{0.8} & \HTEcp{35.1}{1.2} & \HUEcp{207.6}{8.8} & \CcEtc{28.1}{1.0} & \HTEtc{38.0}{1.3} & \HUEtc{203.5}{8.5} \\
\scendash
5\,ms/2\%
  & ML-DSA-65 & \CcCrv{33.0}{1.1} & \HTCrv{42.7}{1.5} & \HUCrv{209.0}{8.8} & \CcEcp{29.6}{1.0} & \HTEcp{43.4}{1.5} & \HUEcp{223.1}{9.4} & \CcEtc{34.1}{1.2} & \HTEtc{46.6}{1.6} & \HUEtc{241.8}{10.2} \\
\scendash
5\,ms/5\%
  & ML-DSA-65 & \CcCrv{47.0}{1.7} & \HTCrv{66.5}{2.4} & \HUCrv{469.7}{19.2} & \CcEcp{44.0}{1.6} & \HTEcp{67.5}{2.4} & \HUEcp{467.7}{19.5} & \CcEtc{43.2}{1.5} & \HTEtc{68.3}{2.4} & \HUEtc{482.5}{19.8} \\
\bottomrule
\end{tabular}%
}
\end{table*}

%% file: sections/06_energy_table.tex
\newcommand{\ebarcell}[3]{%
  \pgfmathparse{(#1)/(#3)*46}\let\bwpt\pgfmathresult%
  \begin{tikzpicture}[x=1pt,y=1pt,baseline=2pt]
    \useasboundingbox (0,-1) rectangle (46,8);
    \draw[cyan!60!black, line width=0.25pt] (0,0) rectangle (46,7);
    \fill[cyan!30!white] (0.2,0.2) rectangle (\bwpt,6.8);
    \node[anchor=east, inner sep=1pt, font=\scriptsize] at (45.5,3.5) {#1($\pm$#2)};
  \end{tikzpicture}%
}

\newcommand{\enabar}{%
  \begin{tikzpicture}[x=1pt,y=1pt,baseline=2pt]
    \useasboundingbox (0,-1) rectangle (46,8);
    \draw[cyan!60!black, line width=0.25pt] (0,0) rectangle (46,7);
    \node[anchor=east, inner sep=1pt, font=\scriptsize] at (45.5,3.5) {---};
  \end{tikzpicture}%
}

\newcommand{\maxECrvC}{992}
\newcommand{\maxECrvHT}{1234}
\newcommand{\maxECrvHU}{7646}

\newcommand{\maxEEcpC}{875}
\newcommand{\maxEEcpHT}{1281}
\newcommand{\maxEEcpHU}{7890}

\newcommand{\maxEEtcC}{975}
\newcommand{\maxEEtcHT}{1373}
\newcommand{\maxEEtcHU}{7748}

\newcommand{\ECcCrv}[2]{\ebarcell{#1}{#2}{\maxECrvC}}
\newcommand{\EHTCrv}[2]{\ebarcell{#1}{#2}{\maxECrvHT}}
\newcommand{\EHUCrv}[2]{\ebarcell{#1}{#2}{\maxECrvHU}}

\newcommand{\ECcEcp}[2]{\ebarcell{#1}{#2}{\maxEEcpC}}
\newcommand{\EHTEcp}[2]{\ebarcell{#1}{#2}{\maxEEcpHT}}
\newcommand{\EHUEcp}[2]{\ebarcell{#1}{#2}{\maxEEcpHU}}

\newcommand{\ECcEtc}[2]{\ebarcell{#1}{#2}{\maxEEtcC}}
\newcommand{\EHTEtc}[2]{\ebarcell{#1}{#2}{\maxEEtcHT}}
\newcommand{\EHUEtc}[2]{\ebarcell{#1}{#2}{\maxEEtcHU}}

\begin{table*}[htbp]
\centering
\caption{Energy consumption (mJ) and standard deviation (inside brackets) across transport modes and algorithm pairs. Bars represent energy normalized to the column maximum.}
\label{tab:energy_results}
\setlength{\tabcolsep}{3pt}
\renewcommand{\arraystretch}{1.1}
\scriptsize
\resizebox{\textwidth}{!}{%
\begin{tabular}{@{}ll
  l@{\hspace{2pt}}l@{\hspace{2pt}}l|
  l@{\hspace{2pt}}l@{\hspace{2pt}}l|
  l@{\hspace{2pt}}l@{\hspace{2pt}}l@{}}
\toprule
& & \multicolumn{3}{c|}{\textbf{Curve25519$\Vert$768}}
  & \multicolumn{3}{c|}{\textbf{ECP-256$\Vert$768}}
  & \multicolumn{3}{c}{\textbf{ECP-384$\Vert$1024}} \\
\textbf{Scenario} & \textbf{Auth}
  & \textbf{Comp} & \textbf{Hyb(TCP)} & \textbf{Hyb(UDP)}
  & \textbf{Comp} & \textbf{Hyb(TCP)} & \textbf{Hyb(UDP)}
  & \textbf{Comp} & \textbf{Hyb(TCP)} & \textbf{Hyb(UDP)} \\
\midrule

\multirow{2}{*}{0\%}
  & ML-DSA-65 & \ECcCrv{384}{2.6} & \EHTCrv{392}{2.6} & \EHUCrv{383}{2.6} & \ECcEcp{387}{2.6} & \EHTEcp{397}{2.6} & \EHUEcp{383}{2.6} & \ECcEtc{462}{2.5} & \EHTEtc{475}{2.5} & \EHUEtc{466}{2.5} \\
  & ML-DSA-87 & \ECcCrv{433}{2.9} & \EHTCrv{445}{3.0} & \EHUCrv{429}{2.9} & \ECcEcp{438}{2.9} & \EHTEcp{452}{3.0} & \EHUEcp{434}{2.9} & \ECcEtc{513}{2.7} & \EHTEtc{534}{2.8} & \EHUEtc{509}{2.7} \\
\scendash
\multirow{2}{*}{0.25\%}
  & ML-DSA-65 & \ECcCrv{401}{5.2} & \EHTCrv{411}{5.0} & \EHUCrv{577}{33.3} & \ECcEcp{398}{5.2} & \EHTEcp{411}{4.8} & \EHUEcp{580}{32.6} & \ECcEtc{476}{5.0} & \EHTEtc{496}{4.9} & \EHUEtc{641}{36.1} \\
  & ML-DSA-87 & \ECcCrv{444}{5.8} & \EHTCrv{456}{5.5} & \EHUCrv{641}{37.0} & \ECcEcp{446}{5.8} & \EHTEcp{460}{5.4} & \EHUEcp{644}{36.2} & \ECcEtc{518}{5.4} & \EHTEtc{542}{5.3} & \EHUEtc{713}{40.2} \\
\scendash
\multirow{2}{*}{0.5\%}
  & ML-DSA-65 & \ECcCrv{407}{7.8} & \EHTCrv{418}{9.0} & \EHUCrv{702}{38.6} & \ECcEcp{393}{7.8} & \EHTEcp{432}{9.6} & \EHUEcp{769}{41.6} & \ECcEtc{477}{9.8} & \EHTEtc{503}{9.3} & \EHUEtc{829}{45.1} \\
  & ML-DSA-87 & \ECcCrv{450}{8.6} & \EHTCrv{466}{10.0} & \EHUCrv{780}{42.9} & \ECcEcp{444}{8.6} & \EHTEcp{469}{10.4} & \EHUEcp{856}{46.3} & \ECcEtc{520}{10.7} & \EHTEtc{547}{10.1} & \EHUEtc{922}{50.1} \\
\scendash
\multirow{2}{*}{1\%}
  & ML-DSA-65 & \ECcCrv{421}{9.3} & \EHTCrv{442}{12.0} & \EHUCrv{1341}{65.7} & \ECcEcp{404}{9.4} & \EHTEcp{442}{12.2} & \EHUEcp{1150}{58.5} & \ECcEtc{478}{11.7} & \EHTEtc{523}{14.0} & \EHUEtc{1526}{75.3} \\
  & ML-DSA-87 & \ECcCrv{475}{10.5} & \EHTCrv{500}{13.5} & \EHUCrv{1491}{73.1} & \ECcEcp{456}{10.6} & \EHTEcp{500}{13.8} & \EHUEcp{1279}{65.1} & \ECcEtc{540}{13.2} & \EHTEtc{591}{15.8} & \EHUEtc{1698}{83.8} \\
\scendash
\multirow{2}{*}{2\%}
  & ML-DSA-65 & \ECcCrv{459}{13.5} & \EHTCrv{505}{16.0} & \EHUCrv{1676}{72.8} & \ECcEcp{434}{13.0} & \EHTEcp{515}{16.4} & \EHUEcp{1769}{78.6} & \ECcEtc{503}{15.2} & \EHTEtc{594}{17.9} & \EHUEtc{1743}{72.7} \\
  & ML-DSA-87 & \ECcCrv{527}{15.5} & \EHTCrv{591}{18.7} & \EHUCrv{1864}{81.0} & \ECcEcp{502}{15.1} & \EHTEcp{601}{19.1} & \EHUEcp{1968}{87.4} & \ECcEtc{574}{17.3} & \EHTEtc{693}{21.5} & \EHUEtc{1937}{80.9} \\
\scendash
\multirow{2}{*}{5\%}
  & ML-DSA-65 & \ECcCrv{607}{19.8} & \EHTCrv{676}{22.4} & \EHUCrv{3620}{148.8} & \ECcEcp{524}{17.3} & \EHTEcp{670}{22.9} & \EHUEcp{3735}{156.7} & \ECcEtc{589}{19.8} & \EHTEtc{767}{24.7} & \EHUEtc{3669}{145.5} \\
  & ML-DSA-87 & \ECcCrv{707}{23.1} & \EHTCrv{856}{28.6} & \EHUCrv{4021}{165.5} & \ECcEcp{628}{20.7} & \EHTEcp{847}{29.8} & \EHUEcp{4149}{174.2} & \ECcEtc{694}{23.4} & \EHTEtc{974}{32.1} & \EHUEtc{3958}{157.0} \\
\scendash
\multirow{2}{*}{10\%}
  & ML-DSA-65 & \ECcCrv{853}{30.5} & \EHTCrv{1032}{39.5} & \EHUCrv{6878}{286.2} & \ECcEcp{739}{25.8} & \EHTEcp{1073}{39.0} & \EHUEcp{7096}{298.7} & \ECcEtc{826}{29.1} & \EHTEtc{1150}{42.5} & \EHUEtc{6971}{293.0} \\
  & ML-DSA-87 & \ECcCrv{992}{35.4} & \EHTCrv{1234}{46.1} & \EHUCrv{7646}{318.5} & \ECcEcp{875}{30.6} & \EHTEcp{1281}{46.5} & \EHUEcp{7890}{332.1} & \ECcEtc{975}{34.4} & \EHTEtc{1373}{50.7} & \EHUEtc{7748}{325.7} \\
\midrule
\multirow{2}{*}{1\,ms/0\%}
  & ML-DSA-65 & \ECcCrv{410}{3.0} & \EHTCrv{433}{2.8} & \EHUCrv{403}{2.8} & \ECcEcp{415}{2.9} & \EHTEcp{438}{2.8} & \EHUEcp{410}{2.8} & \ECcEtc{493}{2.8} & \EHTEtc{517}{2.7} & \EHUEtc{485}{2.7} \\
  & ML-DSA-87 & \ECcCrv{451}{3.4} & \EHTCrv{475}{3.1} & \EHUCrv{469}{3.1} & \ECcEcp{455}{3.3} & \EHTEcp{479}{3.1} & \EHUEcp{476}{3.2} & \ECcEtc{536}{3.1} & \EHTEtc{564}{3.0} & \EHUEtc{558}{3.0} \\
\scendash
\multirow{2}{*}{5\,ms/0\%}
  & ML-DSA-65 & \ECcCrv{522}{4.3} & \EHTCrv{589}{4.0} & \EHUCrv{509}{3.9} & \ECcEcp{526}{4.3} & \EHTEcp{591}{4.0} & \EHUEcp{511}{4.0} & \ECcEtc{603}{4.3} & \EHTEtc{666}{4.0} & \EHUEtc{587}{3.9} \\
  & ML-DSA-87 & \ECcCrv{523}{4.9} & \EHTCrv{591}{4.5} & \EHUCrv{586}{4.4} & \ECcEcp{526}{4.9} & \EHTEcp{596}{4.5} & \EHUEcp{591}{4.5} & \ECcEtc{614}{4.9} & \EHTEtc{685}{4.5} & \EHUEtc{673}{4.4} \\
\scendash
\multirow{2}{*}{10\,ms/0\%}
  & ML-DSA-65 & \ECcCrv{614}{5.9} & \EHTCrv{757}{7.2} & \EHUCrv{745}{7.1} & \ECcEcp{615}{5.8} & \EHTEcp{763}{7.2} & \EHUEcp{748}{7.1} & \ECcEtc{696}{5.9} & \EHTEtc{841}{7.3} & \EHUEtc{836}{7.3} \\
  & ML-DSA-87 & \ECcCrv{695}{6.7} & \EHTCrv{848}{8.1} & \EHUCrv{838}{8.0} & \ECcEcp{697}{6.6} & \EHTEcp{854}{8.1} & \EHUEcp{840}{8.0} & \ECcEtc{782}{6.6} & \EHTEtc{942}{8.2} & \EHUEtc{938}{8.2} \\
\midrule
2\,ms/2\%
  & ML-DSA-87 & \ECcCrv{516}{16.2} & \EHTCrv{601}{20.8} & \EHUCrv{1917}{81.3} & \ECcEcp{495}{16.9} & \EHTEcp{612}{20.9} & \EHUEcp{1823}{77.3} & \ECcEtc{567}{20.2} & \EHTEtc{684}{23.4} & \EHUEtc{1999}{83.5} \\
\scendash
5\,ms/2\%
  & ML-DSA-87 & \ECcCrv{535}{17.8} & \EHTCrv{636}{22.3} & \EHUCrv{799}{33.6} & \ECcEcp{542}{18.3} & \EHTEcp{651}{22.5} & \EHUEcp{1004}{42.3} & \ECcEtc{617}{21.7} & \EHTEtc{707}{24.3} & \EHUEtc{890}{37.5} \\
\scendash
5\,ms/5\%
  & ML-DSA-87 & \ECcCrv{591}{21.4} & \EHTCrv{637}{23.0} & \EHUCrv{1719}{70.3} & \ECcEcp{575}{20.9} & \EHTEcp{625}{22.2} & \EHUEcp{1841}{76.8} & \ECcEtc{618}{21.5} & \EHTEtc{736}{25.9} & \EHUEtc{1867}{76.6} \\
\bottomrule
\end{tabular}%
}
\end{table*}

%% file: sections/06a_ideal.tex

\subsection{Baseline: Ideal Network Conditions}
\label{sec:results:baseline}

The first row of Table~\ref{tab:compvshyb2} corresponds to ideal network conditions, i.e., zero packet loss and
zero network delay. Handshake latencies are within 0.4 ms of each
other for all key-exchange variants for any pairing of crypto primitives and DSA algorithm pairing. Across the entire suite, handshake performance of C/TCP with key-encapsulation crypto primitive Curve25519768||ML-KEM-768 and DSA ML-DSA-44 is the best at average of 11.6 ms and reaching the upper value for  H/UDP with the combination ECP-384||ML-KEM-1024 and ML-DSA-65. Throughput
follows the similar pattern. All three hey-exchange variants deliver within 2--3 connections
of each other spread all key-exchange variants for any pairing of crypto primitives and DSA algorithm pairing. However, this closeness disappears once the packet-loss increases.

\para{Effect of transport choice at 0\% loss.}
The TCP three-way handshake adds 0.1--0.4\,ms for H/TCP relative to H/UDP
on the Curve25519 and ECP-256 pairs.  C/TCP sits 0.3--0.6 ms
above its H/TCP counterpart on the same algorithm pair. Both gaps are
within the standard variations of the measurement and does not indicate transport advantage or penalty for C/TCP. The practical implication
is that, at ideal conditions, the choice between C/TCP, H/TCP, and H/UDP
has no measurable effect on either connection time or throughput.

\para{Effect of classical algorithms.}
The ECP-384 pairs are the only consistent outlier across all three key-variants and all DSA. They run 3.9--4.2 ms slower
than the corresponding Curve25519 pairs and 3.6--3.9 ms slower than the
ECP-256 pairs. This offset is transport-independent, appearing identically
in the C/TCP, H/TCP, and H/UDP columns, and reflects 384-bit field
arithmetic rather than any protocol overhead. ECP-256 and Curve25519 are
not meaningfully separable at this timescale.

\para{DSA overhead:}
As expected, ML-DSA-65 adds more latency and consumes more energy compared to ML-DSA-44 while all other conditions remain the same. 
Stepping from ML-DSA-44 to ML-DSA-65 adds 1.9 ms and 31 mJ. These increments are the same regardless of which key-exchange pair or transport
mode is in use, i.e., a deployment running C/TCP running ECP-384||ML-KEM-1024 with ML-DSA-44
pays the same authentication premium when switching to ML-DSA-65 as a deployment running H/UDP with Cruve25519 and ML-KEM-768 with ML-DSA-44 when switching to ML-DSA-65. An operator migrating from
ML-DSA-44 to full ML-DSA-65 should expect a net cost of approximately
3.0 ms and 30 mJ per handshake, applied as a flat additive term on
whatever baseline their existing deployment shows. This gap can cam more than double when migrating to ML-DSA-87.

\para{Effect on Energy:}
Energy consumptions spans 353 mJ to approximately 511 mJ  per handshake as the DSA and encapsulation crypto suite combination moves from (ML-KEM-44, Curve25519, ML-DSA-44) to (ML-KME-1024, ECP-384, ML-DSA-65). Change of DSA from
ML-DSA-44 to ML-DSA-65 costs roughly 
30 mJ, consistent with the latency increments and independent of
key-exchange variants. At ideal conditions, energy and cycle count are driven almost entirely by the authentication computation. The key exchange contribution is smaller and largely hidden within measurement variance.

The observations under ideal network conditions establish that the performance is impacted by the DSA used and not the underlying transport. However, the impact of DSA in ideal conditions in marginal. As we see next, this observations changes drastically under non-deal conditions. 

%% file: sections/06b_under_loss.tex

\subsection{Effect of Packet Loss}
\label{sec:results:loss}

\begin{figure*}[htb]
\centering
\includegraphics[width=\textwidth]{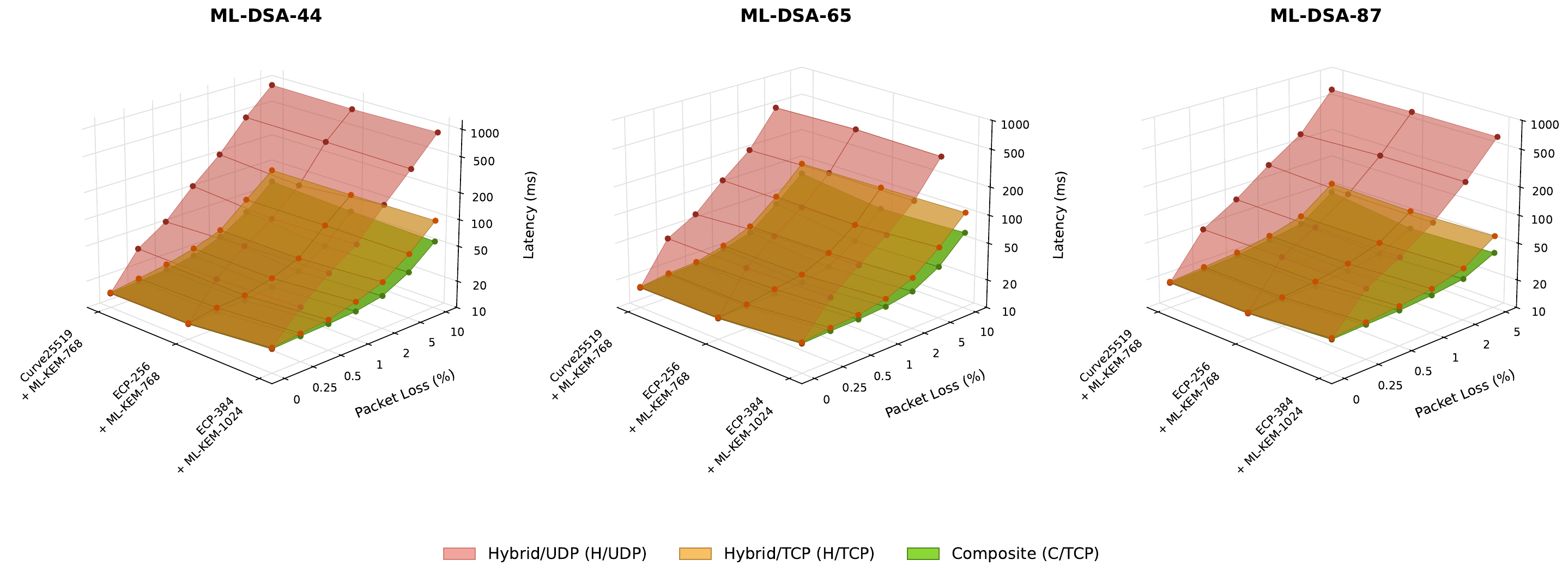}
\Description{Three 3D surface plots showing handshake latency under packet loss for ML-DSA-44, 65, and 87. Hybrid/UDP diverges exponentially from the TCP variants above 0.25 percent loss.}
\caption{Handshake latency performance under difference crypto-primitives and packet loss. Each sub-figure corresponds to different DSA-- ML-DSA-44 (left),
  ML-DSA-65 (center), and ML-DSA-87 (right). Each sub-figure shows performance of different key-exchage varaints -- C/TCP (Green), H/TCP (Orange), H/UDP (Red).}
\label{fig:loss}
\vspace{-15pt}
\end{figure*}

Figure~\ref{fig:loss} shows the latency surfaces across seven loss rates for
all three key-exchange variant with ML-DSA-44, ML-DSA-65, and ML-DSA-87. No confidence bars are shown to avoid cluster. Even with the confidence bars, the separation is non-overlapping for packet loss of more than 0.5\%. 
All key-exchange variants are within a few milliseconds of handshake latency near
at 0\% packet loss. As the packet loss increases beyond 0.25\%, H/UDP diverges sharply while C/TCP and H/TCP continue along
a gradual slope. 

At 0.25\% packet loss, a rate that indoor 802.11n links routinely exceed,
Crv25519$+$ML-KEM-768\,(H/UDP) with ML-DSA-44 reaches 36.4\,ms while
Crv25519$\Vert$ML-KEM-768\,(C/TCP) completes in 15.3\,ms, a factor of
2.4$\times$. By 1\% loss the same pair reads 118.4\,ms for H/UDP against
18.2\,ms for C/TCP, an improvement of 6.5$\times$. At 10\% loss, H/UDP reaches
918 ms for the Curve25519 and ML-KEM-768 pair with ML-DSA-65, while C/TCP for the same
combination finishes in 67.2\,ms, an improvement by a factor of roughly 13.7$\times$.
The absolute gap exceeds 850 ms at that point.

The throughput column (see Appendix) tells the same story in a different unit. At 0\%
packet loss, all three key-exchange at least 31 connections per second (conn/s) 
At 10\% loss, H/UDP with ML-DSA-65 falls to 0.9 conn/s, effectively one
successful handshake per second, while C/TCP still achieves 4.0 conn/s
and H/TCP 4.1 conn/s. Throughput under H/UDP collapses faster than latency
grows because retransmission timeouts serialise the handshake pipeline. A gateway blocked on a 1.0s IKEv2 timer cannot accept new connections
during that wait.

Between C/TCP and H/TCP, the dominant cost is round-trip count rather than
payload size. C/TCP completes one fewer round trip than
H/TCP, and at 10\% loss that difference jumps to 25--35 ms in
Table~\ref{tab:compvshyb2}: C/TCP takes an average in 43--58 ms across all three
key-exchange variants and both DSAs, while H/TCP runs in 82--87 ms.
ECP-256$\Vert$ML-KEM-768 (C/TCP) with ML-DSA-44 achieves the lowest absolute latency at 10\% loss among all Composite pairs at 43.3 ms.

Comparing the three sub figures in Figure~\ref{fig:loss} directly, the vertical
offset between the ML-DSA-44, ML-DSA-65, and ML-DSA-87 surfaces is
1.9--4.1\,ms. The slope with which H/UDP separates from the TCP variants reads the same
across all three panels at every measured loss rate. From the data, the
the signature scheme shifts the latency surface up by a fixed amount, though
it does not change how quickly H/UDP diverges from C/TCP as loss increases.

%% file: sections/06c_rto_analysis.tex

\subsection{Retransmission Timeout Analysis}
\label{sec:results:rto}

The latency explosion in H/UDP observed in Section~\ref{sec:results:loss}
is not a smooth degradation, it is a step function driven by a discrete
protocol event. In UDP mode, retransmissions are governed entirely at the
application layer by the IKE daemon. When a UDP fragment carrying the ML-KEM
ciphertext is dropped, the IKEv2 state machine receives no response and waits
for its application-level timer to fire. The duration of the $n$-th retry is
calculated by strongSwan as $T_n = \text{timeout} \times \text{base}^{(n-1)}$.
Our benchmark uses an aggressive configuration (\texttt{retransmit\_timeout = 1.0},
\texttt{retransmit\_base = 1.2}, and \texttt{retransmit\_tries = 3}). Consequently,
a single lost UDP fragment delays the entire handshake by a minimum of
$1.0 \times 1.2^0 = 1.0$\,s, regardless of the underlying round-trip time.

TCP retransmissions, by contrast, operate below the IKE state machine at the
Linux kernel level and are invisible to strongSwan's \texttt{charon} daemon.
The kernel manages its own Retransmission Timeout (RTO) intervals using
system parameters like \texttt{net.ipv4.tcp\_rto\_min\_us} (set to 200\,ms).
The kernel performs rapid, segment-level exponential backoff on dropped packets,
allowing C/TCP and H/TCP to recover from loss in hundreds of milliseconds without
ever triggering the heavy 1.0\,s application-level IKEv2 timer.

\begin{figure*}[htb]
\centering
\includegraphics[width=\textwidth]{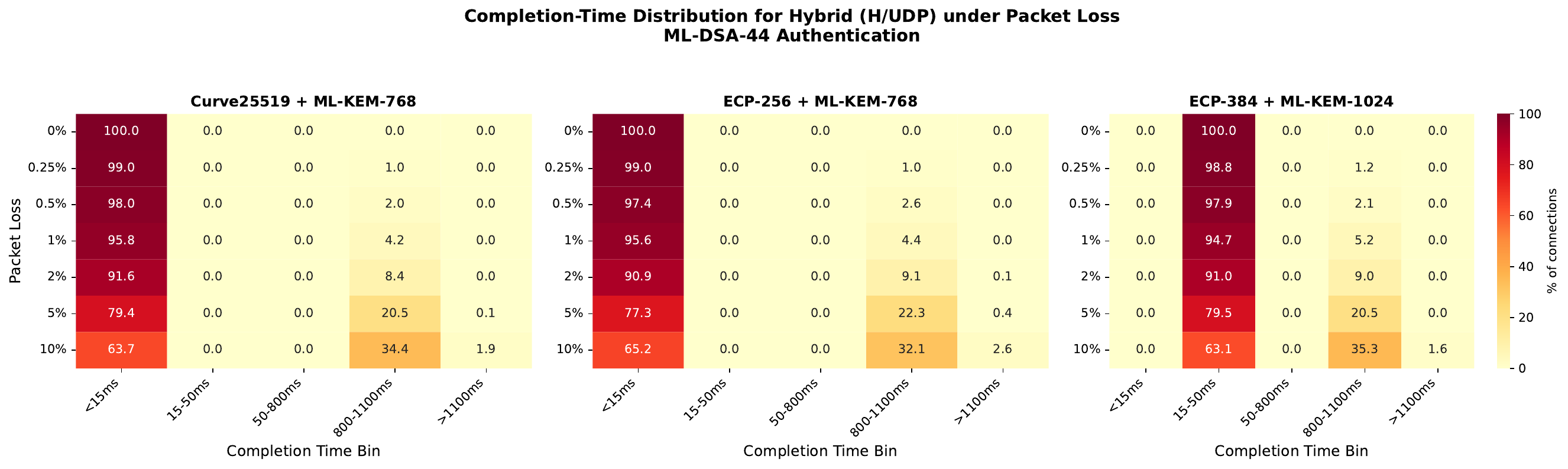}
\Description{Grid of three heatmaps showing completion time distribution for Hybrid/UDP configurations under ML-DSA-44 and packet loss. Most connections shift to the 800-1100ms bin above 1 percent loss.}
\caption{Per-handshake completion-time distribution for Hybrid\,(H/UDP)
  under increasing packet loss with ML-DSA-44. Cells represent the percentage of handshakes completing within a given time bin.}
\label{fig:rto_heatmap}
\end{figure*}

Figure~\ref{fig:rto_heatmap} makes this distribution explicit for H/UDP under ML-DSA-44. It presents heatmaps on cells formed by handshake time and packet loss. Each cell displays the exact fraction of H/UDP handshakes falling into a specific completion-time and loss rates.

At 0\% and 0.25\% loss, over 90\% of handshakes complete in the fastest
bin ($<$15\,ms), and the tail bins are nearly empty. From 1\% loss onward,
mass migrates from the fast bin into the 800--1{,}100\,ms bin, exactly
the width of one IKEv2 retransmission timeout. By 10\% loss, the 800--1{,}100\,ms
bin contains 74--76\% of all handshakes across the algorithm pairs. The fast bin has been almost entirely
emptied. Put differently, at 10\% packet loss, three out of every four
H/UDP VPN tunnels are waiting a full second before completing.

The three panels in Figure~\ref{fig:rto_heatmap} are remarkably similar in shape.
This is structurally expected as the timeout is triggered entirely by the
fragmentation of the preceding key exchange message. Because the ML-KEM payloads impose
equivalent fragmentation penalties, they produce nearly identical timeout collapse curves.
While Figure~\ref{fig:rto_heatmap} isolates ML-DSA-44, our measurements confirm that this exact pattern holds across all signature schemes (including RSA-2048 and ML-DSA-65). This occurs because the signature payload is transmitted only after the vulnerable \textsc{ike\_intermediate} exchange has completed. A corresponding discrete-bin distribution is not shown for the TCP variants because TCP recovers from dropped packets transparently in the kernel and does not experience application-level timeout clustering. This results in the smooth, lower-bound latency surfaces seen previously in Figure~\ref{fig:loss}.

\para{Throughput collapse.}
The timeout event does more than add latency to a single handshake. Because
the IKEv2 daemon serialises retransmissions within each SA negotiation,
a gateway experiencing 10\% packet loss on H/UDP is blocked for approximately
0.75\,s per handshake on average. This is why the throughput at 10\% loss
for H/UDP drops to 0.9--1.5\,conn/s across all three algorithm pairs,
substantially below what the processor or the network would limit it to
under ideal conditions. C/TCP, which absorbs the same physical packet loss
inside the TCP stack without exposing it to IKEv2, maintains 4.0--5.1\,conn/s
at the same loss rate.

%% file: sections/06d_under_delay.tex

\subsection{Effect of Network Delay}
\label{sec:results:delay}

We first fix the packet loss at 10\% to isolate the effect of network delay.  Table~\ref{tab:compvshyb2} isolates this effect in the rows corresponding to 1\,ms, 5\,ms, and 10\,ms network delay. As packet delivery is guaranteed in this case, UDP fragment reassembly succeeds reliably on the first attempt, preventing the application-level timeouts detailed in Section~\ref{sec:results:rto}. Consequently, H/UDP and H/TCP converge, leaving round-trip count as the sole differentiator. At 10 ms delay, H/TCP with  Curvev25519, ML-KEM-768, and ML-DSA-44 combination requires 41.8 ms to complete, while the C/TCP
takes 31.4 ms under the same combination. This 10.4 ms latency gap represents the exact cost of the single additional network round trip imposed by the sequential key exchange design of Hybrid mode.

Network delay impacts throughput not by blocking daemons on timeouts, but simply by occupying connection slots for longer durations. For the Curve25519, ML-KEM-768, and ML-DSA-44 combination, throughput drops from 38.9 conn/s under ideal conditions down to 15.2 conn/s at 10 ms delay for C/TCP. For the same scenario,  H/TCP drops slightly faster to 13.1 conn/s due to its extra round trip. Crucially, pure network delay induces no additional computational work. CPU cycle counts and energy consumption remain statistically identical to those recorded under ideal conditions, isolating performance degradation purely to network flight time. Finally, the fixed additive cost of heavier signature schemes remains preserved: shifting from ML-DSA-44 to ML-DSA-65 adds 2.7 ms to both C/TCP and H/TCP handshakes at 10 ms delay. The choice of signature scheme shifts the latency baseline, but the proportional delay penalty remains identical across all cryptographic combinations.

%% file: sections/06e_combined.tex

\subsection{Effect of Combined Delay and Packet Loss}
\label{sec:results:combined}

Real WAN and mobile links impose both delay and packet loss simultaneously. We collected a complete matrix of combined scenarios covering delays form 1 ms to 5 ms and packet loss rate from 0.5\% to 5.0\%. The two network impairments act on entirely separate protocol mechanisms. While loss stalls retransmission timers, the delay multiplies the cost of each additional round trip. Consequently, the two impairments do not interact. The performance advantage of C/TCP over H/TCP grows predictably with both delay and loss independently.

\begin{figure*}[t]
\centering
\includegraphics[width=\textwidth]{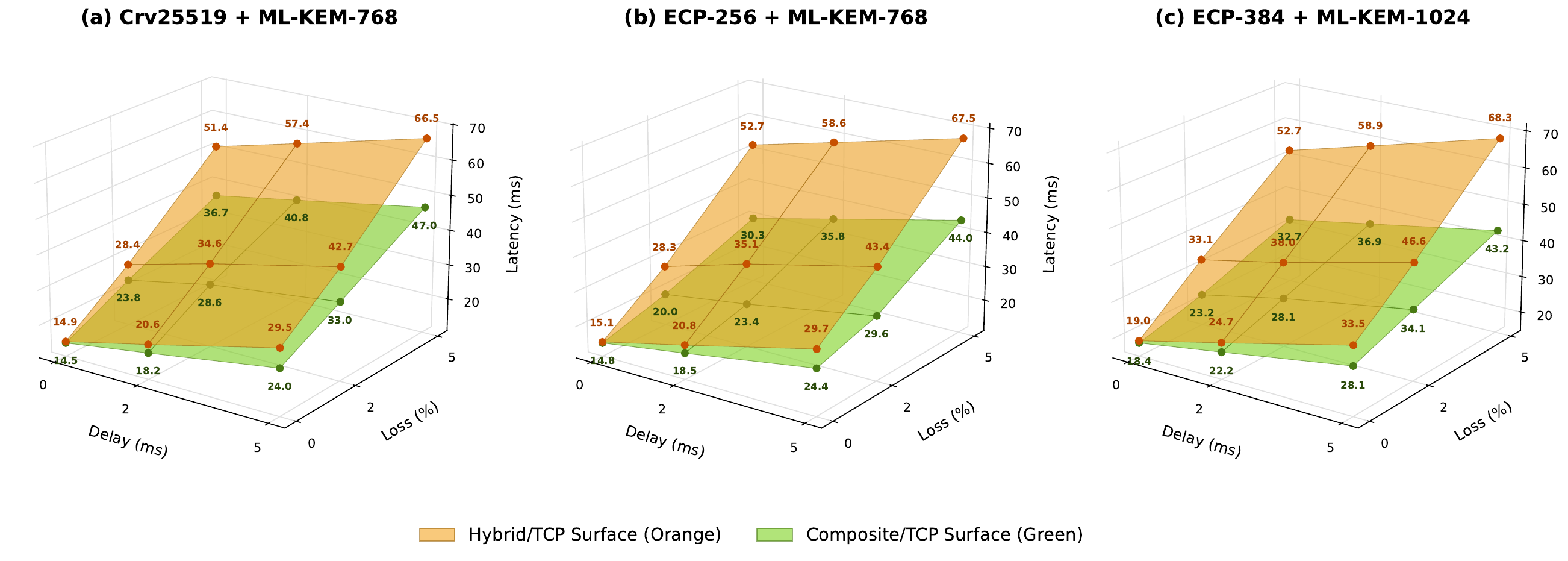}
\Description{Three 3D surface plots showing latency for C/TCP (green) and H/TCP (orange) under combinations of 1 to 5 ms delay and 0.5 to 5 percent loss. The orange surface is consistently above the green surface.}
\caption{Hanshake latency under combined delay and loss (ML-DSA-44). \textcolor{compositecol}{\rule{0.8em}{0.5em}} C/TCP \textcolor{hybridcol!80!black}{\rule{0.8em}{0.5em}} H/TCP.}
\label{fig:combined3d}
\vspace{-15pt}
\end{figure*}

Fig.~\ref{fig:combined3d} presents this independent interaction as three 3D latency surfaces, one per each encapsulation crypto-primitives with ML-DSA-44. 
H/UDP is excluded from these surfaces because its exponential timeout-driven collapse (200--900\,ms) would compress the TCP surfaces into visually indistinguishable flat planes. The vertical gap between the C/TCP and H/TCP surfaces represents the round-trip premium, which scales up additively along the delay axis and multiplicatively along the loss axis. Again, the plots are shown without confidence intervals to avoid clutter. 

Selected cross-sections from this grid are detailed in Table~\ref{tab:compvshyb2}. At 2 ms delay and 2\% loss, a profile typical of conservative enterprise WAN links, Composite\,(C/TCP) with ML-DSA-44 finishes in 21.0--24.2\,ms while Hybrid\,(H/TCP) requires 30.3--35.4\,ms across the three key exchange pairs. This translates to a 28--50\% latency penalty on every connection setup. Under a stressed mobile scenario of 5\,ms delay and 5\% loss, the performance gap widens significantly where composite completes in 36.3--44.5 ms while Hybrid requires 59.6--70.3 ms. The data confirms that under real-world conditions combining moderate delay with measurable packet drop, the architectural efficiency of Composite transport delivers substantial resilience.

%% file: sections/06g_energy.tex

\subsection{Energy and CPU Consumption}
\label{sec:results:energy}

\begin{figure*}[ht]
\centering
\includegraphics[width=\textwidth]{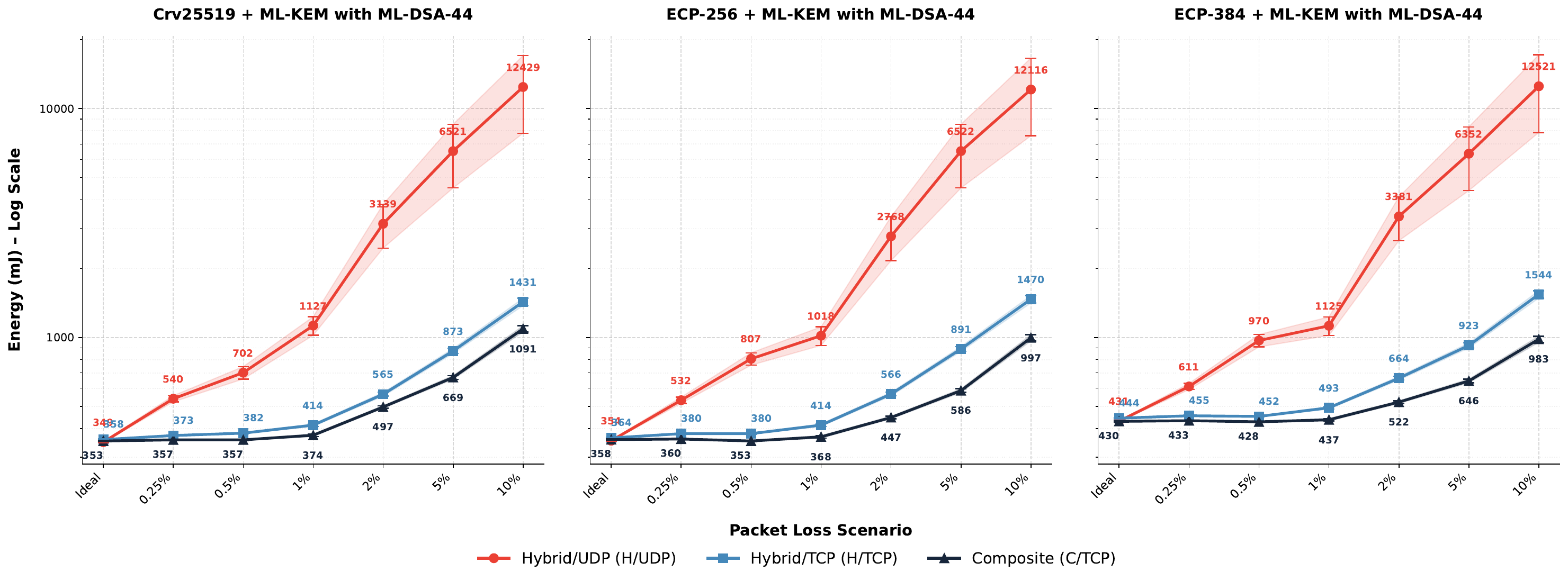}
\Description{Line charts showing energy consumption scaling exponentially with packet loss for Hybrid/UDP, while TCP modes remain stable and low.}
\caption{Energy scaling under packet loss (log scale). H/UDP rises exponentially due to application timeouts, while C/TCP remains stable via kernel recovery.}
\label{fig:energy}
\vspace{-10pt}
\end{figure*}

Figure~\ref{fig:energy} presents energy per handshake (mJ) across the loss spectrum. We can observe a massive divergence above 1\% packet drop rate. Table~\ref{tab:energy_results} provides the corresponding exact measurements alongside CPU cycle counts in Table~\ref{tab:extended_results_cpu_cycles}. 
Notice that, under ideal conditions, computational footprints remain tightly clustered regardless of transport mode. 
Further, the baseline CPU execution ranges between 46--68 million cycles, with the corresponding energy consumption ranging between 349--475\,mJ across all configurations.

The operationally significant finding is the severe computational inversion that occurs under network impairment. As packet loss increases, application-level IKEv2 timeouts (Section~\ref{sec:results:rto}) force the CPU to remain active during retransmission polling. At 5\% loss, the per-handshake energy consumption of H/UDP increases to 3,620--6,522 mJ, which is 5.9--11.1$\times$ higher than C/TCP. Further, at 10\% loss, H/UDP incurs up to 3.48 billion CPU cycles per handshake, driving energy consumption to an extreme 6,878--12,521 mJ. In contrast, C/TCP absorbs the impairment within the kernel, capping execution at 190--255 million cycles and limiting energy draw to 739--1,091 mJ. H/TCP fares better than H/UDP but consistently worse than C/TCP, penalized by the energy drawn during its additional network round trip.

While upgrading the signature scheme from ML-DSA-44 to ML-DSA-65 adds a  $\sim$30 mJ overhead under ideal conditions, this signature overhead is entirely eclipsed by transport inefficiency under packet loss. As seen at 10\% loss, the random distribution of multi-second application timeouts causes energy profiles to fluctuate wildly, rendering signature baseline offsets operationally irrelevant when the transport layer collapses.

%% file: sections/06h_auth_comparison.tex

\subsection{Effect of Authentication Algorithm}
\label{sec:results:auth}

\begin{figure*}[t]
\centering
\includegraphics[width=\textwidth]{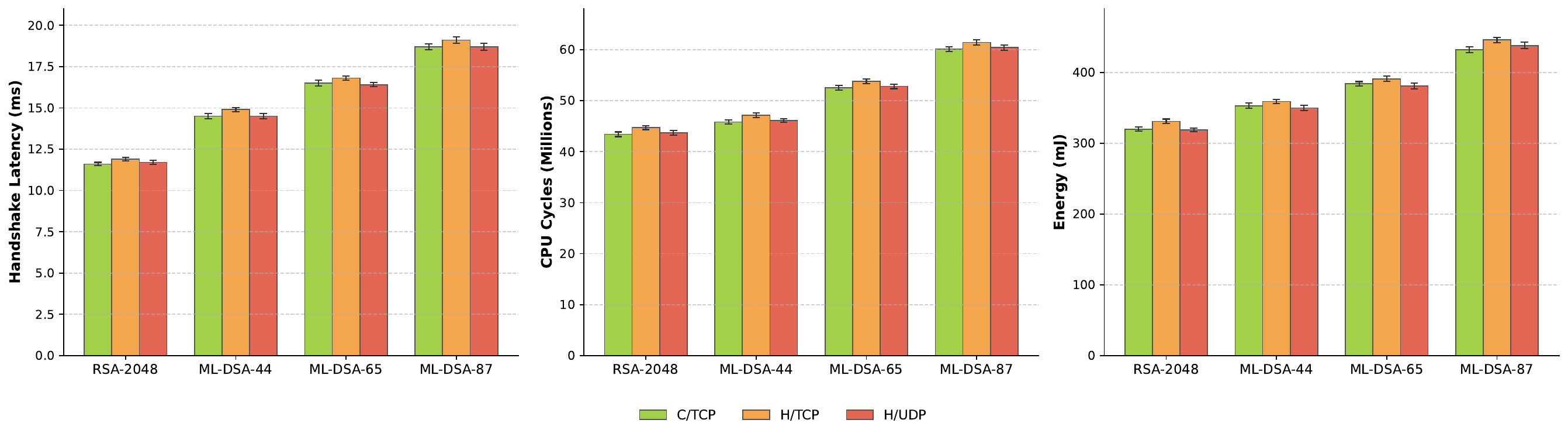}
\Description{Bar charts comparing latency, CPU cycles, and energy across RSA-2048 and three ML-DSA variants. All metrics show a stepped increase as security level rises.}
\vspace{-12pt}
\caption{Signature Algorithm (ideal conditions). Transitioning from RSA to ML-DSA introduces progressive, additive penalties across all metrics.}
\label{fig:auth_overhead}
\vspace{-12pt}
\end{figure*}

We now focus on evaluating the intrinsic computational overhead of the signature scheme.
Figure~\ref{fig:auth_overhead} presents the performance degradation incurred when migrating from the classical RSA-2048 baseline to the three NIST-standardized post-quantum signature schemes: ML-DSA-44 (category 2), ML-DSA-65 (category 3), and ML-DSA-87 (category 5). In order to eliminate transport variance, we strictly examine the Curve25519 key exchange paired with Composite\,(C/TCP) under ideal network conditions (0\,ms delay, 0\% loss).

Under ideal conditions, the cryptographic execution bottleneck reveals a strict degradation across all four primary metrics as security levels scale. The classical RSA-2048 baseline establishes the performance ceiling, completing handshakes in 11.6 ms, sustaining a throughput of 38.9 conn/s, and consuming 318.8 mJ. Transitioning to the ML-DSA-44 DSA increases connection time to 14.5 ms, reduces throughput to 36.2 conn/s, and rises energy consumption to 352.8 mJ. Upgrading to higher security parameters introduces strictly additive penalties. ML-DSA-65 pushes latency to 16.5 ms (34.0 conn/s) and energy to 384.1 mJ (52.6 million CPU cycles). Finally, the category-5 ML-DSA-87 maximizes the algorithmic burden, completing handshakes in 18.6 ms 31.5 (conn/s) and driving energy consumption to 433.0 mJ (60.0 million CPU cycles).

%% file: sections/07_wire_bytes.tex

\subsection{IKE Wire Bytes}
\label{sec:wire}

A preliminary assessment might indicate that C/TCP can transmit more data overall, since they pack both classical and post-quantum keys into a single \textsc{ike\_sa\_init} payload. However, our wire-level measurements show that this is not the case and H/TCP and H/UDP actually transmits more total bytes per handshake.
Although the Hybrid approach benefits from a much smaller initial message, it requires a separate \textsc{ike\_intermediate} exchange to transmit the post-quantum keys. This extra round trip adds unavoidable framing, encapsulation, and TCP acknowledgement overhead. 
To quantify this, we measured the total bytes exchanged over the network for the combination of Curve25519, ML-KEM-768, and ML-DSA-44. C/TCP transmits a total of 19,123 bytes for the complete handshake. For the same combination, hybrid variants expand this size to 19,688 bytes. This pattern holds across all security levels. For the heavy pair of ECP-384 and ML-KEM-1024), C/TCP requires 20,064 bytes in total, while the Hybrid variant requires 20,875 bytes. 

Across all evaluated crypto suites, the extra \textsc{ike\_intermediate} exchange introduces roughly 565 to 811 bytes of additional overhead into the Hybrid handshake. Consequently, the C/TCP proves to be more byte-efficient in practice, which complements its performance advantages under poor networks conditions in the previous subsections.

%% file: sections/08_related.tex
\section{Related Work}
\label{sec:related}
Recent work on post-quantum IKEv2 and IPsec shows that the main systems challenge is not the standalone cost of new primitives, but the interaction between larger control-plane messages, additional exchange rounds, and impaired network conditions. Bae et al.~\cite{bae2023} established that integrating post-quantum key exchange into strongSwan substantially increases IKEv2 packet size and can push exchanges beyond the fragmentation threshold. Mutlugun et al.~\cite{mutlugun2024} extended that observation to satellite settings, showing that fragmentation and loss can dominate handshake latency, although high propagation delay can partially mask the effect. Twardokus et al.~\cite{twardokus2025} reached a similar conclusion under Wi-Fi and Internet-scale conditions, reporting severe setup overhead under modest loss and arguing that current RFC-based designs remain fragile on impaired links. Perera et al.~\cite{perera2026} and Iliadis-Apostolidis et al.~\cite{qrons2025} showed that ML-KEM- and Dilithium-based IPsec are practical in 5G and high-throughput environments, but neither examined lossy handshake behavior in detail.

The closest prior work to our transport comparison is Blanco-Romero et al.~\cite{blancoromerohybrid2025}, who compared sequential RFC~9370 hybrids in IKEv2 protocol and showed that the additional \textsc{ike\_intermediate} round trip is a structural latency bottleneck. Taken together, this literature shows that PQC stresses IKEv2 primarily through packetization and transport rather than computation alone. What remains unexamined is whether moving only the IKE control plane to TCP, while keeping ESP on UDP, can eliminate fragmentation costs without incurring TCP-over-TCP pathologies. Also, this work did not consider the composite mechanism. The primary focus remained on the fragmentation issues caused by the quantum crypto payload and their serial distribution. They propose parallel hybrid
approaches to eliminate multiplicative latency
penalties inherent in sequential methods without any transport switch.  

The fragmentation problem itself predates post-quantum migration. Kaufman et al.~\cite{kaufman2003} and Zhou et al.~\cite{zhou2010} identified the reliance of UDP-based IKE on IP reassembly and motivated application-layer fragmentation mechanisms later standardized in RFC~7383~\cite{rfc7383}. Smyslov~\cite{smyslov2018compact} likewise sought to reduce IKEv2 message size to avoid fragmentation. More recently, He et al.~\cite{he2025} showed that post-quantum IKEv2 over fragmented \textsc{ike\_intermediate} exchanges suffers heavily from full-message retransmission under loss and proposed selective retransmission as a mitigation. That approach improves the fragmented design, but it does not remove the underlying dependence on large UDP-borne control messages. Our approach is complementary in motivation but different in mechanism: rather than optimizing retransmission after fragmentation occurs, it removes fragmentation exposure from the control plane. TCP encapsulation appears in the standards literature, but only as a fallback mode that carries both IKE and ESP over TCP. RFC~9329~\cite{rfc9329} defines that mechanism, and Smyslov~\cite{smyslov2019} provides implementation guidance without performance evaluation. Prior work has treated this design cautiously because the VPN literature has long documented the TCP-in-TCP meltdown problem~\cite{tcpmeltdown,coonjah2018,harkanson2019}. This concern also explains why Twardokus et al.~\cite{twardokus2025} did not evaluate TCP transport despite identifying fragmentation as a central problem. To our knowledge, no prior study has evaluated TCP encapsulation specifically for post-quantum IKEv2 handshakes, and no prior implementation has demonstrated the split-transport design of IKE over TCP with ESP over UDP enabled by \textsc{separate\_transports} ~\cite{reddydraft}.

Compared with key exchange, post-quantum authentication remains less well characterized in IKEv2. Existing IPsec studies include Falcon, Dilithium, or ML-DSA in end-to-end evaluations~\cite{mutlugun2024,qrons2025}, but they do not isolate signature overhead across transport choices and network conditions. Our results complement that literature by showing that authentication overhead is additive and largely transport-independent, which suggests that key-exchange and authentication migration can be reasoned about separately.

Energy is less explored still. Prior methodology papers validate Intel RAPL as a practical basis for server-side energy measurement~\cite{rapl,raffin2024}, and adjacent protocol studies use it to examine communication costs under loss~\cite{zatuchin2025,schoeffel2022,halak2024}. However, no prior IPsec or IKEv2 study, to our knowledge, measures per-handshake energy across transport choices under impaired network conditions. This is the final gap addressed by our study.

%% file: sections/09_conclusion.tex
\section{Conclusion}
\label{sec:conclusion}

In this work we evaluated the performance of IPSec with NIST-standardized encapsulation and digital signature algorithms ~\cite{fips203,fips204}. We studied three key-sharing mechanisms in the IKEv2 protocol that deliver the classical and post-quantum crypto payloads differently using different transport protocols, namely Hybrid key sharing via UDP (H/UDP), Hybrid key sharing via TCP (H/TCP) and Composite key sharing via TCP (C/TCP). 


We implemented all the three key-sharing variants in strongSwan and evaluated their performance against various metrics like handshake latency, throughput, energy consumption, CPU cycles under various network conditions. Our analysis demonstrated that C/TCP performs better then both H/UDP and H/TCP under all network condition and its performance relative to others only get better as the network conditions worsen. In the addition to performance, it offer multiple benefits as discussed below.  


\begin{enumerate}[label=(\arabic*), leftmargin=2em]
\item \textbf{Security:} 
Unlike the H/TCP and H/UDP, C/TCP does not need keys derived from the classical algorithms to encrypts post-quantum crypto payloads. TCP handles the reliable transport of all the crypto payload without requiring any encryption. This advantage makes it a strong candidate for PQ-only systems where classical algorithmics are completely deprecated.   

\item \textbf{Deployability:} 
TCP for transport in IKEv2 is already proposed in  RFC~9329~\cite{rfc9329}. However,  it is not yet widely deployed across the operational ecosystem due to the TCP meltdown phenomenon \cite{tcpmeltdown}. It is mostly used as a fall-back mechanism when UDP is blocked in the network. C/TCP using TCP for key-share in IKEv2 and UDP in ESP through transport switch, demonstrated that the larger data transfer can still continue to happen on UDP without TCP meltdown as in the current deployment, only key-share can be done using TCP to ensure reliable transfer of large quantum crypto payloads. Our results provide strong evidence that this transport switch offers improved performance across all network conditions. We believe our evaluation results will pay ways of wider adaption of TCP in IKEv2 for sharing the quantum crypto primitives. 


\item \textbf{Cost vs security performance:}
Both Hybrid UDP and Hybird TCP key sharing mechanisms introduces a third protocol round trip through the mandatory \textsc{ike\_intermediate} (ADDKE) exchange, and our results show this is the single greatest source of measurable performance loss. At 10\% packet loss over TCP, this one additional exchange costs 25 to 39 ms per handshake, representing 30 to 47\% higher latency compared to C/TCP. Also, the overhead in H/TCP and H/UDP is not marginal under degraded network conditions. At 10\% loss, 74 to 76\% of all H/UDP encounter 1.0s retransmission timeout, causing exponential latency collapse. C/TCP is also more byte-efficient despite carrying a substantially larger \textsc{ike\_sa\_init}: eliminating the ADDKE exchange and its acknowledgement reduces the aggregate wire footprint by 582 to 827 bytes per handshake. 
A particularly salient finding concerns ML-DSA-65. Its authentication overhead of 4.6 ms per handshake, combined with its increased signature and verification cost relative to ML-DSA-44, produces a disproportionately unfavourable cost-to-security ratio. ML-DSA-44 achieves full FIPS~204 Level~2 compliance at 2.6 ms and, based on our measurements, appears to offer the more favourable cost-to-security ratio for most deployment contexts outside those with explicit Level~3 requirements.
\end{enumerate}

\para{Limitations.}
All measurements reported here were conducted within a single-host loopback testbed using emulated network impairment. Cross-architecture reproducibility and validation on live Internet paths remain essential open work, in keeping with the acknowledged scope boundaries of prior studies.
\newpage

%% file: sections/appendix.tex
\newpage

\appendix
\section*{APPENDIX}

Table~\ref{tab:extended_results} reports handshake throughput (conn/s) for every scenario in Table~\ref{tab:compvshyb2}. At ideal conditions, all three transports are within 1\,conn/s of each other: C/TCP peaks at 38.9\,conn/s (Crv25519$\Vert$768, ML-DSA-44) while H/TCP and H/UDP reach 38.3 and 39.1\,conn/s respectively; the ECP-384$\Vert$1024 pair runs 4--5\,conn/s lower across all modes. A key observation is that H/UDP throughput collapses \emph{faster} than its latency grows: at 10\% loss, H/UDP falls to 0.9--1.5\,conn/s (roughly one tunnel per second) because the 1.0\,s IKEv2 retransmission timer serialises the entire connection pipeline (Section~\ref{sec:results:rto}), while C/TCP sustains 4.0--5.4\,conn/s and H/TCP 4.1--4.5\,conn/s by absorbing loss inside the kernel. Under pure delay, H/UDP and H/TCP converge---at 10\,ms both reach 12--15\,conn/s---because no retransmission fires; under combined 5\,ms/5\% stress, H/UDP falls to 1.7--2.7\,conn/s against C/TCP's 7.2--8.2\,conn/s. Stepping ML-DSA-44 to ML-DSA-65 costs a consistent 2--3\,conn/s regardless of scenario, confirming the transport-independence of signature overhead.

\newcommand{\tbarcell}[3]{%
  \pgfmathparse{(#2)/(#3)*36}\let\bwpt\pgfmathresult%
  \begin{tikzpicture}[x=1pt,y=1pt,baseline=2pt]
    \useasboundingbox (0,-1) rectangle (36,8);
    \draw[cyan!60!black, line width=0.25pt] (0,0) rectangle (36,7);
    \fill[cyan!30!white] (0.2,0.2) rectangle (\bwpt,6.8);
    \node[anchor=east, inner sep=1pt, font=\scriptsize] at (35.5,3.5) {#2};
  \end{tikzpicture}%
}

\newcommand{\tnabar}{%
  \begin{tikzpicture}[x=1pt,y=1pt,baseline=2pt]
    \useasboundingbox (0,-1) rectangle (36,8);
    \draw[cyan!60!black, line width=0.25pt] (0,0) rectangle (36,7);
    \node[anchor=east, inner sep=1pt, font=\scriptsize] at (35.5,3.5) {---};
  \end{tikzpicture}%
}

\newcommand{\maxTCrvC}{38.9}
\newcommand{\maxTCrvHT}{38.3}
\newcommand{\maxTCrvHU}{39.1}

\newcommand{\maxTEcpC}{38.6}
\newcommand{\maxTEcpHT}{37.9}
\newcommand{\maxTEcpHU}{38.7}

\newcommand{\maxTEtcC}{33.9}
\newcommand{\maxTEtcHT}{33.1}
\newcommand{\maxTEtcHU}{33.8}

\newcommand{\TCcCrv}[2]{\tbarcell{#1}{#2}{\maxTCrvC}}
\newcommand{\THTCrv}[2]{\tbarcell{#1}{#2}{\maxTCrvHT}}
\newcommand{\THUCrv}[2]{\tbarcell{#1}{#2}{\maxTCrvHU}}

\newcommand{\TCcEcp}[2]{\tbarcell{#1}{#2}{\maxTEcpC}}
\newcommand{\THTEcp}[2]{\tbarcell{#1}{#2}{\maxTEcpHT}}
\newcommand{\THUEcp}[2]{\tbarcell{#1}{#2}{\maxTEcpHU}}

\newcommand{\TCcEtc}[2]{\tbarcell{#1}{#2}{\maxTEtcC}}
\newcommand{\THTEtc}[2]{\tbarcell{#1}{#2}{\maxTEtcHT}}
\newcommand{\THUEtc}[2]{\tbarcell{#1}{#2}{\maxTEtcHU}}

\newcommand{\tscendash}{\noalign{\vskip1pt}\cdashline{1-11}[0.5pt/1.5pt]\noalign{\vskip1pt}}

\newcommand{\cbarcell}[3]{%
  \pgfmathparse{(#1)/(#3)*36}\let\bwpt\pgfmathresult%
  \begin{tikzpicture}[x=1pt,y=1pt,baseline=2pt]
    \useasboundingbox (0,-1) rectangle (36,8);
    \draw[cyan!60!black, line width=0.25pt] (0,0) rectangle (36,7);
    \fill[cyan!30!white] (0.2,0.2) rectangle (\bwpt,6.8);
    \node[anchor=east, inner sep=1pt, font=\scriptsize] at (35.5,3.5) {#1};
  \end{tikzpicture}%
}

\newcommand{\maxCCrvC}{255}
\newcommand{\maxCCrvHT}{331}
\newcommand{\maxCCrvHU}{3295}

\newcommand{\maxCEcpC}{202}
\newcommand{\maxCEcpHT}{344}
\newcommand{\maxCEcpHU}{3486}

\newcommand{\maxCEtcC}{219}
\newcommand{\maxCEtcHT}{361}
\newcommand{\maxCEtcHU}{3357}

\newcommand{\CCcCrv}[2]{\cbarcell{#1}{#2}{\maxCCrvC}}
\newcommand{\CHTCrv}[2]{\cbarcell{#1}{#2}{\maxCCrvHT}}
\newcommand{\CHUCrv}[2]{\cbarcell{#1}{#2}{\maxCCrvHU}}

\newcommand{\CCcEcp}[2]{\cbarcell{#1}{#2}{\maxCEcpC}}
\newcommand{\CHTEcp}[2]{\cbarcell{#1}{#2}{\maxCEcpHT}}
\newcommand{\CHUEcp}[2]{\cbarcell{#1}{#2}{\maxCEcpHU}}

\newcommand{\CCcEtc}[2]{\cbarcell{#1}{#2}{\maxCEtcC}}
\newcommand{\CHTEtc}[2]{\cbarcell{#1}{#2}{\maxCEtcHT}}
\newcommand{\CHUEtc}[2]{\cbarcell{#1}{#2}{\maxCEtcHU}}

\begin{table*}[htbp]
\centering
\caption{Extended Performance Results: (a) Throughput.}
\label{tab:extended_results}

\vspace{5pt}
\textbf{(a) Throughput (conn/s)} \\
\vspace{2pt}
\resizebox{\textwidth}{!}{%
\begin{tabular}{@{}ll
  l@{\hspace{2pt}}l@{\hspace{2pt}}l|
  l@{\hspace{2pt}}l@{\hspace{2pt}}l|
  l@{\hspace{2pt}}l@{\hspace{2pt}}l@{}}
\toprule
& & \multicolumn{3}{c|}{\textbf{Curve25519$\Vert$768}}
  & \multicolumn{3}{c|}{\textbf{ECP-256$\Vert$768}}
  & \multicolumn{3}{c}{\textbf{ECP-384$\Vert$1024}} \\
\textbf{Scenario} & \textbf{Auth}
  & \textbf{Comp} & \textbf{Hyb(TCP)} & \textbf{Hyb(UDP)}
  & \textbf{Comp} & \textbf{Hyb(TCP)} & \textbf{Hyb(UDP)}
  & \textbf{Comp} & \textbf{Hyb(TCP)} & \textbf{Hyb(UDP)} \\
\midrule
\multirow{2}{*}{0\%}
  & ML-DSA-44  & \TCcCrv{0.0}{38.9} & \THTCrv{0.0}{38.3} & \THUCrv{0.0}{39.1} & \TCcEcp{0.0}{38.6} & \THTEcp{0.0}{37.9} & \THUEcp{0.0}{38.7} & \TCcEtc{0.0}{33.9} & \THTEtc{0.0}{33.1} & \THUEtc{0.0}{33.8} \\
  & ML-DSA-65 & \TCcCrv{0.0}{36.2} & \THTCrv{0.0}{35.8} & \THUCrv{0.0}{36.4} & \TCcEcp{0.0}{35.9} & \THTEcp{0.0}{35.4} & \THUEcp{0.0}{36.0} & \TCcEtc{0.0}{31.8} & \THTEtc{0.0}{31.2} & \THUEtc{0.0}{31.8} \\
\tscendash
\multirow{2}{*}{0.25\%}
  & ML-DSA-44  & \TCcCrv{0.0}{32.5} & \THTCrv{0.0}{32.0} & \THUCrv{0.0}{25.9} & \TCcEcp{0.0}{34.1} & \THTEcp{0.0}{33.1} & \THUEcp{0.0}{25.9} & \TCcEtc{0.0}{29.8} & \THTEtc{0.0}{28.6} & \THUEtc{0.0}{22.5} \\
  & ML-DSA-65 & \TCcCrv{0.0}{32.2} & \THTCrv{0.0}{31.0} & \THUCrv{0.0}{19.5} & \TCcEcp{0.0}{30.8} & \THTEcp{0.0}{29.0} & \THUEcp{0.0}{19.6} & \TCcEtc{0.0}{28.2} & \THTEtc{0.0}{27.2} & \THUEtc{0.0}{18.9} \\
\tscendash
\multirow{2}{*}{0.5\%}
  & ML-DSA-44  & \TCcCrv{0.0}{29.5} & \THTCrv{0.0}{29.1} & \THUCrv{0.0}{19.5} & \TCcEcp{0.0}{30.3} & \THTEcp{0.0}{28.9} & \THUEcp{0.0}{17.7} & \TCcEtc{0.0}{27.2} & \THTEtc{0.0}{26.1} & \THUEtc{0.0}{17.1} \\
  & ML-DSA-65 & \TCcCrv{0.0}{28.8} & \THTCrv{0.0}{26.6} & \THUCrv{0.0}{13.5} & \TCcEcp{0.0}{26.8} & \THTEcp{0.0}{26.5} & \THUEcp{0.0}{12.1} & \TCcEtc{0.0}{26.2} & \THTEtc{0.0}{24.2} & \THUEtc{0.0}{11.3} \\
\tscendash
\multirow{2}{*}{1\%}
  & ML-DSA-44  & \TCcCrv{0.0}{24.6} & \THTCrv{0.0}{21.6} & \THUCrv{0.0}{12.1} & \TCcEcp{0.0}{24.3} & \THTEcp{0.0}{21.5} & \THUEcp{0.0}{11.9} & \TCcEtc{0.0}{21.9} & \THTEtc{0.0}{21.7} & \THUEtc{0.0}{10.7} \\
  & ML-DSA-65 & \TCcCrv{0.0}{22.1} & \THTCrv{0.0}{19.4} & \THUCrv{0.0}{7.2} & \TCcEcp{0.0}{24.3} & \THTEcp{0.0}{21.2} & \THUEcp{0.0}{8.1} & \TCcEtc{0.0}{21.6} & \THTEtc{0.0}{18.8} & \THUEtc{0.0}{7.6} \\
\tscendash
\multirow{2}{*}{2\%}
  & ML-DSA-44  & \TCcCrv{0.0}{15.7} & \THTCrv{0.0}{16.0} & \THUCrv{0.0}{7.1} & \TCcEcp{0.0}{17.0} & \THTEcp{0.0}{15.4} & \THUEcp{0.0}{6.5} & \TCcEtc{0.0}{16.3} & \THTEtc{0.0}{14.6} & \THUEtc{0.0}{6.7} \\
  & ML-DSA-65 & \TCcCrv{0.0}{16.1} & \THTCrv{0.0}{14.1} & \THUCrv{0.0}{4.0} & \TCcEcp{0.0}{16.9} & \THTEcp{0.0}{14.6} & \THUEcp{0.0}{4.6} & \TCcEtc{0.0}{15.1} & \THTEtc{0.0}{13.4} & \THUEtc{0.0}{3.8} \\
\tscendash
\multirow{2}{*}{5\%}
  & ML-DSA-44  & \TCcCrv{0.0}{8.5} & \THTCrv{0.0}{7.8} & \THUCrv{0.0}{3.2} & \TCcEcp{0.0}{8.7} & \THTEcp{0.0}{7.8} & \THUEcp{0.0}{2.7} & \TCcEtc{0.0}{8.6} & \THTEtc{0.0}{8.4} & \THUEtc{0.0}{2.9} \\
  & ML-DSA-65 & \TCcCrv{0.0}{8.2} & \THTCrv{0.0}{7.9} & \THUCrv{0.0}{1.8} & \TCcEcp{0.0}{9.1} & \THTEcp{0.0}{7.9} & \THUEcp{0.0}{1.9} & \TCcEtc{0.0}{9.3} & \THTEtc{0.0}{7.4} & \THUEtc{0.0}{1.9} \\
\tscendash
\multirow{2}{*}{10\%}
  & ML-DSA-44  & \TCcCrv{0.0}{4.8} & \THTCrv{0.0}{4.5} & \THUCrv{0.0}{1.4} & \TCcEcp{0.0}{5.4} & \THTEcp{0.0}{4.2} & \THUEcp{0.0}{1.5} & \TCcEtc{0.0}{4.9} & \THTEtc{0.0}{4.2} & \THUEtc{0.0}{1.4} \\
  & ML-DSA-65 & \TCcCrv{0.0}{4.0} & \THTCrv{0.0}{4.1} & \THUCrv{0.0}{1.0} & \TCcEcp{0.0}{4.8} & \THTEcp{0.0}{4.1} & \THUEcp{0.0}{1.0} & \TCcEtc{0.0}{5.1} & \THTEtc{0.0}{4.3} & \THUEtc{0.0}{0.9} \\
\midrule
\multirow{2}{*}{1\,ms/0\%}
  & ML-DSA-44  & \TCcCrv{0.0}{33.7} & \THTCrv{0.0}{32.1} & \THUCrv{0.0}{33.7} & \TCcEcp{0.0}{33.4} & \THTEcp{0.0}{31.8} & \THUEcp{0.0}{33.4} & \TCcEtc{0.0}{29.8} & \THTEtc{0.0}{28.4} & \THUEtc{0.0}{29.7} \\
  & ML-DSA-65 & \TCcCrv{0.0}{31.4} & \THTCrv{0.0}{30.0} & \THUCrv{0.0}{31.5} & \TCcEcp{0.0}{31.0} & \THTEcp{0.0}{29.9} & \THUEcp{0.0}{31.1} & \TCcEtc{0.0}{27.9} & \THTEtc{0.0}{26.7} & \THUEtc{0.0}{27.9} \\
\tscendash
\multirow{2}{*}{5\,ms/0\%}
  & ML-DSA-44  & \TCcCrv{0.0}{21.9} & \THTCrv{0.0}{19.6} & \THUCrv{0.0}{21.9} & \TCcEcp{0.0}{21.9} & \THTEcp{0.0}{19.4} & \THUEcp{0.0}{21.8} & \TCcEtc{0.0}{20.3} & \THTEtc{0.0}{18.1} & \THUEtc{0.0}{20.1} \\
  & ML-DSA-65 & \TCcCrv{0.0}{21.2} & \THTCrv{0.0}{19.0} & \THUCrv{0.0}{21.2} & \TCcEcp{0.0}{21.1} & \THTEcp{0.0}{18.9} & \THUEcp{0.0}{21.1} & \TCcEtc{0.0}{19.6} & \THTEtc{0.0}{17.6} & \THUEtc{0.0}{19.5} \\
\tscendash
\multirow{2}{*}{10\,ms/0\%}
  & ML-DSA-44  & \TCcCrv{0.0}{15.2} & \THTCrv{0.0}{13.1} & \THUCrv{0.0}{15.2} & \TCcEcp{0.0}{15.1} & \THTEcp{0.0}{13.0} & \THUEcp{0.0}{15.1} & \TCcEtc{0.0}{14.3} & \THTEtc{0.0}{12.4} & \THUEtc{0.0}{14.3} \\
  & ML-DSA-65 & \TCcCrv{0.0}{14.8} & \THTCrv{0.0}{12.8} & \THUCrv{0.0}{14.8} & \TCcEcp{0.0}{14.7} & \THTEcp{0.0}{12.8} & \THUEcp{0.0}{14.7} & \TCcEtc{0.0}{13.9} & \THTEtc{0.0}{12.1} & \THUEtc{0.0}{13.9} \\
\midrule
\multirow{2}{*}{2\,ms/2\%}
  & ML-DSA-44  & \TCcCrv{0.0}{14.2} & \THTCrv{0.0}{13.7} & \THUCrv{0.0}{6.4} & \TCcEcp{0.0}{16.0} & \THTEcp{0.0}{12.9} & \THUEcp{0.0}{6.2} & \TCcEtc{0.0}{15.1} & \THTEtc{0.0}{13.1} & \THUEtc{0.0}{5.7} \\
  & ML-DSA-65 & \TCcCrv{0.0}{14.4} & \THTCrv{0.0}{13.1} & \THUCrv{0.0}{4.4} & \TCcEcp{0.0}{14.9} & \THTEcp{0.0}{12.3} & \THUEcp{0.0}{4.3} & \TCcEtc{0.0}{13.3} & \THTEtc{0.0}{12.8} & \THUEtc{0.0}{4.1} \\
\tscendash
\multirow{2}{*}{5\,ms/2\%}
  & ML-DSA-44  & \TCcCrv{0.0}{12.3} & \THTCrv{0.0}{10.6} & \THUCrv{0.0}{6.2} & \TCcEcp{0.0}{13.2} & \THTEcp{0.0}{11.9} & \THUEcp{0.0}{5.8} & \TCcEtc{0.0}{12.8} & \THTEtc{0.0}{10.6} & \THUEtc{0.0}{5.7} \\
  & ML-DSA-65 & \TCcCrv{0.0}{12.0} & \THTCrv{0.0}{10.4} & \THUCrv{0.0}{3.9} & \TCcEcp{0.0}{12.4} & \THTEcp{0.0}{10.5} & \THUEcp{0.0}{3.8} & \TCcEtc{0.0}{11.6} & \THTEtc{0.0}{10.6} & \THUEtc{0.0}{3.6} \\
\tscendash
\multirow{2}{*}{5\,ms/5\%}
  & ML-DSA-44  & \TCcCrv{0.0}{7.4} & \THTCrv{0.0}{6.5} & \THUCrv{0.0}{2.7} & \TCcEcp{0.0}{8.2} & \THTEcp{0.0}{6.9} & \THUEcp{0.0}{3.0} & \TCcEtc{0.0}{7.2} & \THTEtc{0.0}{6.5} & \THUEtc{0.0}{2.9} \\
  & ML-DSA-65 & \TCcCrv{0.0}{7.5} & \THTCrv{0.0}{6.4} & \THUCrv{0.0}{1.8} & \TCcEcp{0.0}{8.1} & \THTEcp{0.0}{6.1} & \THUEcp{0.0}{1.9} & \TCcEtc{0.0}{7.2} & \THTEtc{0.0}{6.0} & \THUEtc{0.0}{1.7} \\
\bottomrule
\end{tabular}%
}

\end{table*}

Table~\ref{tab:extended_results_cpu_cycles} reports per-handshake CPU cycle counts (millions), a hardware-level metric independent of scheduler jitter and network timing. At ideal conditions, cycles range from 46\,M (Crv25519$\Vert$768, C/TCP, ML-DSA-44) to 66\,M (ECP-384$\Vert$1024, C/TCP, ML-DSA-65); the 6--7\,M increment per DSA level appears identically across all three transports, confirming signature computation is fully decoupled from transport overhead. Under packet loss, H/UDP cycles explode due to IKEv2 retransmission polling (Section~\ref{sec:results:rto}): at 10\% loss H/UDP peaks at 3,041--3,486\,M cycles ($\approx$3.0--3.5\,billion) while C/TCP is bounded by 190--255\,M and H/TCP by 309--361\,M---directly confirming the figures cited in Section~\ref{sec:results:energy}. Under pure delay, all transports rise proportionally with round-trip count rather than exhibiting a step collapse; under 5\,ms/5\% combined stress, H/UDP surges to 1,601--1,855\,M while C/TCP remains at 144--178\,M, confirming loss as the dominant overhead driver.

\begin{table*}[htbp]
\centering
\caption{ Performance Results: CPU Cycles (millions).}
\label{tab:extended_results_cpu_cycles}

\vspace{2pt}
\resizebox{\textwidth}{!}{%
\begin{tabular}{@{}ll
  l@{\hspace{2pt}}l@{\hspace{2pt}}l|
  l@{\hspace{2pt}}l@{\hspace{2pt}}l|
  l@{\hspace{2pt}}l@{\hspace{2pt}}l@{}}
\toprule
& & \multicolumn{3}{c|}{\textbf{Curve25519$\Vert$768}}
  & \multicolumn{3}{c|}{\textbf{ECP-256$\Vert$768}}
  & \multicolumn{3}{c}{\textbf{ECP-384$\Vert$1024}} \\
\textbf{Scenario} & \textbf{Auth}
  & \textbf{Comp} & \textbf{Hyb(TCP)} & \textbf{Hyb(UDP)}
  & \textbf{Comp} & \textbf{Hyb(TCP)} & \textbf{Hyb(UDP)}
  & \textbf{Comp} & \textbf{Hyb(TCP)} & \textbf{Hyb(UDP)} \\
\midrule

\multirow{2}{*}{0\%}
  & ML-DSA-44 & \CCcCrv{46}{0.0} & \CHTCrv{47}{0.0} & \CHUCrv{46}{0.0} & \CCcEcp{47}{0.0} & \CHTEcp{48}{0.0} & \CHUEcp{47}{0.0} & \CCcEtc{59}{0.0} & \CHTEtc{61}{0.0} & \CHUEtc{60}{0.0} \\
  & ML-DSA-65 & \CCcCrv{53}{0.0} & \CHTCrv{54}{0.0} & \CHUCrv{53}{0.0} & \CCcEcp{53}{0.0} & \CHTEcp{55}{0.0} & \CHUEcp{54}{0.0} & \CCcEtc{66}{0.0} & \CHTEtc{68}{0.0} & \CHUEtc{66}{0.0} \\
\scendash
\multirow{2}{*}{0.25\%}
  & ML-DSA-44 & \CCcCrv{48}{0.0} & \CHTCrv{53}{0.0} & \CHUCrv{121}{0.0} & \CCcEcp{49}{0.0} & \CHTEcp{55}{0.0} & \CHUEcp{118}{0.0} & \CCcEtc{61}{0.0} & \CHTEtc{66}{0.0} & \CHUEtc{130}{0.0} \\
  & ML-DSA-65 & \CCcCrv{57}{0.0} & \CHTCrv{59}{0.0} & \CHUCrv{148}{0.0} & \CCcEcp{55}{0.0} & \CHTEcp{58}{0.0} & \CHUEcp{148}{0.0} & \CCcEtc{68}{0.0} & \CHTEtc{73}{0.0} & \CHUEtc{155}{0.0} \\
\scendash
\multirow{2}{*}{2\%}
  & ML-DSA-44 & \CCcCrv{77}{0.0} & \CHTCrv{93}{0.0} & \CHUCrv{744}{0.0} & \CCcEcp{65}{0.0} & \CHTEcp{93}{0.0} & \CHUEcp{650}{0.0} & \CCcEtc{75}{0.0} & \CHTEtc{109}{0.0} & \CHUEtc{799}{0.0} \\
  & ML-DSA-65 & \CCcCrv{83}{0.0} & \CHTCrv{99}{0.0} & \CHUCrv{723}{0.0} & \CCcEcp{71}{0.0} & \CHTEcp{102}{0.0} & \CHUEcp{768}{0.0} & \CCcEtc{81}{0.0} & \CHTEtc{115}{0.0} & \CHUEtc{760}{0.0} \\
\scendash
\multirow{2}{*}{5\%}
  & ML-DSA-44 & \CCcCrv{122}{0.0} & \CHTCrv{172}{0.0} & \CHUCrv{1618}{0.0} & \CCcEcp{100}{0.0} & \CHTEcp{176}{0.0} & \CHUEcp{1609}{0.0} & \CCcEtc{108}{0.0} & \CHTEtc{176}{0.0} & \CHUEtc{1565}{0.0} \\
  & ML-DSA-65 & \CCcCrv{143}{0.0} & \CHTCrv{173}{0.0} & \CHUCrv{1734}{0.0} & \CCcEcp{109}{0.0} & \CHTEcp{166}{0.0} & \CHUEcp{1835}{0.0} & \CCcEtc{118}{0.0} & \CHTEtc{193}{0.0} & \CHUEtc{1767}{0.0} \\
\scendash
\multirow{2}{*}{10\%}
  & ML-DSA-44 & \CCcCrv{226}{0.0} & \CHTCrv{309}{0.0} & \CHUCrv{3133}{0.0} & \CCcEcp{202}{0.0} & \CHTEcp{316}{0.0} & \CHUEcp{3041}{0.0} & \CCcEtc{190}{0.0} & \CHTEtc{329}{0.0} & \CHUEtc{3151}{0.0} \\
  & ML-DSA-65 & \CCcCrv{255}{0.0} & \CHTCrv{331}{0.0} & \CHUCrv{3295}{0.0} & \CCcEcp{199}{0.0} & \CHTEcp{344}{0.0} & \CHUEcp{3486}{0.0} & \CCcEtc{219}{0.0} & \CHTEtc{361}{0.0} & \CHUEtc{3357}{0.0} \\
\midrule
\multirow{2}{*}{1\,ms/0\%}
  & ML-DSA-44 & \CCcCrv{53}{0.0} & \CHTCrv{58}{0.0} & \CHUCrv{57}{0.0} & \CCcEcp{54}{0.0} & \CHTEcp{58}{0.0} & \CHUEcp{58}{0.0} & \CCcEtc{67}{0.0} & \CHTEtc{72}{0.0} & \CHUEtc{71}{0.0} \\
  & ML-DSA-65 & \CCcCrv{59}{0.0} & \CHTCrv{64}{0.0} & \CHUCrv{63}{0.0} & \CCcEcp{60}{0.0} & \CHTEcp{65}{0.0} & \CHUEcp{64}{0.0} & \CCcEtc{73}{0.0} & \CHTEtc{78}{0.0} & \CHUEtc{77}{0.0} \\
\scendash
\multirow{2}{*}{5\,ms/0\%}
  & ML-DSA-44 & \CCcCrv{78}{0.0} & \CHTCrv{97}{0.0} & \CHUCrv{97}{0.0} & \CCcEcp{80}{0.0} & \CHTEcp{98}{0.0} & \CHUEcp{98}{0.0} & \CCcEtc{92}{0.0} & \CHTEtc{111}{0.0} & \CHUEtc{110}{0.0} \\
  & ML-DSA-65 & \CCcCrv{86}{0.0} & \CHTCrv{104}{0.0} & \CHUCrv{104}{0.0} & \CCcEcp{87}{0.0} & \CHTEcp{105}{0.0} & \CHUEcp{105}{0.0} & \CCcEtc{100}{0.0} & \CHTEtc{118}{0.0} & \CHUEtc{118}{0.0} \\
\scendash
\multirow{2}{*}{10\,ms/0\%}
  & ML-DSA-44 & \CCcCrv{113}{0.0} & \CHTCrv{148}{0.0} & \CHUCrv{148}{0.0} & \CCcEcp{114}{0.0} & \CHTEcp{149}{0.0} & \CHUEcp{149}{0.0} & \CCcEtc{126}{0.0} & \CHTEtc{163}{0.0} & \CHUEtc{162}{0.0} \\
  & ML-DSA-65 & \CCcCrv{119}{0.0} & \CHTCrv{154}{0.0} & \CHUCrv{155}{0.0} & \CCcEcp{121}{0.0} & \CHTEcp{155}{0.0} & \CHUEcp{156}{0.0} & \CCcEtc{134}{0.0} & \CHTEtc{168}{0.0} & \CHUEtc{170}{0.0} \\
\midrule
\multirow{2}{*}{2\,ms/2\%}
  & ML-DSA-44 & \CCcCrv{94}{0.0} & \CHTCrv{115}{0.0} & \CHUCrv{657}{0.0} & \CCcEcp{76}{0.0} & \CHTEcp{116}{0.0} & \CHUEcp{706}{0.0} & \CCcEtc{92}{0.0} & \CHTEtc{126}{0.0} & \CHUEtc{692}{0.0} \\
  & ML-DSA-65 & \CCcCrv{97}{0.0} & \CHTCrv{124}{0.0} & \CHUCrv{815}{0.0} & \CCcEcp{87}{0.0} & \CHTEcp{126}{0.0} & \CHUEcp{760}{0.0} & \CCcEtc{99}{0.0} & \CHTEtc{136}{0.0} & \CHUEtc{804}{0.0} \\
\scendash
\multirow{2}{*}{5\,ms/2\%}
  & ML-DSA-44 & \CCcCrv{109}{0.0} & \CHTCrv{142}{0.0} & \CHUCrv{711}{0.0} & \CCcEcp{98}{0.0} & \CHTEcp{145}{0.0} & \CHUEcp{759}{0.0} & \CCcEtc{113}{0.0} & \CHTEtc{156}{0.0} & \CHUEtc{823}{0.0} \\
  & ML-DSA-65 & \CCcCrv{115}{0.0} & \CHTCrv{148}{0.0} & \CHUCrv{690}{0.0} & \CCcEcp{104}{0.0} & \CHTEcp{154}{0.0} & \CHUEcp{890}{0.0} & \CCcEtc{119}{0.0} & \CHTEtc{162}{0.0} & \CHUEtc{782}{0.0} \\
\scendash
\multirow{2}{*}{5\,ms/5\%}
  & ML-DSA-44 & \CCcCrv{157}{0.0} & \CHTCrv{223}{0.0} & \CHUCrv{1601}{0.0} & \CCcEcp{147}{0.0} & \CHTEcp{227}{0.0} & \CHUEcp{1595}{0.0} & \CCcEtc{144}{0.0} & \CHTEtc{230}{0.0} & \CHUEtc{1645}{0.0} \\
  & ML-DSA-65 & \CCcCrv{178}{0.0} & \CHTCrv{224}{0.0} & \CHUCrv{1714}{0.0} & \CCcEcp{156}{0.0} & \CHTEcp{217}{0.0} & \CHUEcp{1818}{0.0} & \CCcEtc{154}{0.0} & \CHTEtc{247}{0.0} & \CHUEtc{1855}{0.0} \\
\bottomrule
\end{tabular}%
}
\vspace{-10pt}
\end{table*}